\documentclass[11pt,reqno]{amsart}
\pdfoutput=1

\usepackage{amssymb}
\usepackage{microtype}
\usepackage{mathtools}
\usepackage{multirow}

\usepackage[margin=1.25in, marginparwidth=1in]{geometry}

\usepackage{hyperref}
\hypersetup{
  pdftitle={Dispersive Fractalization in Linear and Nonlinear Fermi-Pasta-Ulam-Tsingou Lattices},
  pdfauthor={Peter J. Olver and Ari Stern},
  bookmarksopen=false,
}

\input otex.sty

\def\i{\,\ro i\,}
\def\O#1{\ro O({#1})} \def\Oh#1{\O{h^{#1}}}
\def\om{\mu} \def\cm{c} \def\m{m} \def\M{M}

\begin{document}

\title[Dispersive Fractalization in Fermi--Pasta--Ulam--Tsingou Lattices]{Dispersive Fractalization in Linear and Nonlinear Fermi--Pasta--Ulam--Tsingou Lattices}

\author{Peter J.~Olver}
\address{School of Mathematics,
University of Minnesota,
Minneapolis, MN\quad 55455}
\email{olver@umn.edu}

\author{Ari Stern}
\address{Department of Mathematics and Statistics, Washington University in St.~Louis}
\email{stern@wustl.edu}

\date{\today}

\begin{abstract} We investigate, both analytically and numerically, dispersive fractalization and quantization of solutions to periodic linear and nonlinear Fermi--Pasta--Ulam--Tsingou  systems. When subject to periodic boundary conditions and discontinuous initial conditions, \eg a step function, both the linearized and nonlinear continuum models for FPUT exhibit fractal solution profiles at irrational times (as determined by the coefficients and the length of the interval) and quantized profiles (piecewise constant or perturbations thereof) at rational times.  We observe a similar effect in the linearized FPUT chain at times $t$ where these models have validity, namely $t = \O{h^{-2}}$, where $h$ is proportional to the intermass spacing or, equivalently, the reciprocal of the number of masses.  For nonlinear periodic FPUT systems, our numerical results suggest a somewhat similar behavior in the presence of small nonlinearities, which disappears as the nonlinear force increases in magnitude. However, these phenomena are manifested on very long time intervals, posing a severe challenge for numerical integration as the number of masses increases.  Even with the high-order splitting methods used here, our numerical investigations are limited to nonlinear FPUT chains with a smaller number of masses than would be needed to resolve this question unambiguously.
\end{abstract}

\maketitle

\vskip15pt

{\it Our problem turned out to have been felicitously chosen.  The results were entirely different qualitatively from what even Fermi, with his great knowledge of wave motions, had expected.  \dots\ To our surprise, the string started playing a game of musical chairs, only between several low notes, and perhaps even more amazingly, after what would have been several hundred ordinary up and down vibrations, it came back almost exactly to its original sinusoidal shape.}

\vskip.1in
\hfill --- \ Stanislaw Ulam, \rf{Ulam; pp. 226--7} \hglue .4in\
\vskip.1in

\vskip5pt

\Section 1 Introduction and Historical Perspective.

The early 1950s witnessed the birth of the world's first all purpose electronic computers\fnote{The term ``computer'' originally refereed to a human, usually female, who performed calculations using mechanical and, later, electro-mechanical calculating machines. One example was the first author's mother, Grace E.~Olver, n\'ee Smith.  Later, following their eventual disappearance from the workforce --- some became the first computer programmers --- what were initially termed  ``electronic computers'' became what we now call ``computers''.}, thereby bringing hitherto infeasible numerical calculations  into the realm of possibility.  With the Los Alamos MANIAC electronic computer at their disposal, Enrico Fermi, John Pasta, and Stanislaw Ulam introduced a simple one-dimensional system, consisting of  a chain of masses connected by springs with nonlinear restoring forces, forming a simplified model for crystals evolving towards thermal equilibrium.  They called upon Mary Tsingou, a former human computer, to write a program (a fairly complex undertaking necessitated by the limited capabilities of MANIAC) to automatically perform the required numerical calculations.  What we now call the Fermi--Pasta--Ulam--Tsingou\fnote{The system is traditionally referred to as the Fermi--Pasta--Ulam (FPU) problem, in recognition of the authors of \rf{FPU}.
However, the first page of the report states ``Report written by Fermi, Pasta, and Ulam.  Work done by Fermi, Pasta, Ulam, and Tsingou.'' In more recent authorship conventions, she would have been listed as a coauthor.
Under her married name Mary Tsingou Menzel, she was a coauthor of an important 1972 paper \rf{TuckMenzel} that describes further computational work on the system performed in 1961. See \rf{FPUT} for additional historical details.} (FPUT) problem is rightly celebrated as one of the very first electronic computer experiments, and certainly the first that produced novel behavior.    The surprise was that the FPUT dynamics, at least on moderately long time intervals, did not proceed to thermalization, \ie exhibit ergodicity, as expected, but rather exhibited an unexpected recurrence, in which energy from the low frequency modes would initially spread out into some of the higher modes but, after a certain time period, the system would almost entirely return to its initial configuration, as eloquently described in the above quote from Ulam's autobiography. It can be argued that the FPUT calculation ``sparked a revolution in modern science'' by inaugurating the contemporary fields of \is{computational physics}, \rf{DPR}, and \is{experimental mathematics}, meaning ``computer-based investigations designed to give insight into complex mathematical and physical problems that are inaccessible, at least initially, using more traditional forms of analysis'', \rf{PZHC}.

In an attempt to understand this intriguing and unexpected phenomenon, Zabusky and Kruskal, \rf{ZabKru}, proposed a continuum model\fnote{Roughly speaking, a continuum model arises as the number of masses goes to infinity. However, the model's derivation and regime of validity is a little subtle; see below for details.} that, in its unidirectional manifestation, turned out to be the Korteweg--deVries (KdV) equation, originally derived by Boussinesq, \rf{Boussinesq}, in his pioneering studies of surface water waves.    Zabusky and Kruskal's numerical integration of the periodic initial-boundary value problem for the Korteweg--deVries equation, starting with a smooth initial profile, led to their discovery of the soliton\fnote{Although the ``solitons'' they observed were in fact finite gap solutions composed of cnoidal waves, \rf{DMN,McKvM}. True solitons only arise when the equation is posed on the entire real line.} and the consequent creation of an entirely new branch of mathematics ---  integrable nonlinear partial differential equations --- whose remarkable repercussions continue to this day, \rf{DJ}.  The overall impact of the FPUT numerical experiment on modern mathematics and physics cannot be understated.

Much of the subsequent extensive research into the FPUT problem has concentrated on understanding their original surprising observation of the non-ergodic (almost) recurrence of the initial state; see \rf{Ford} for historical remarks up to 1992.  One key issue is to determine whether thermalization or ergodicity occurs if one extends the time range to be sufficiently long. An initial conjecture was that, while on a relatively short time scale the system returned close to its original configuration, subsequent behavior at each such ``period'' would take it farther and farther away.  This was disproved in \rf{TuckMenzel}, where Tuck and  Menzel (Tsingou) observed ``superperiods'' after which the initially increasing deviations decrease to a much smaller value.  On the other hand, for sufficiently large nonlinearities, the FPUT recurrence no longer exists, and the systems exhibits a ``strong stochasticity threshold'', \rf{LPRSV1,LPRSV2}.  See \rf{Weissert} and \rf{Ford} for reviews.

 Explanations of the observed behavior have also appealed to Kolmogorov--Arnold--Moser (KAM) theory, \rf{KAMP, Weissert}, in which one views FPUT as a perturbation of either the linearized system, or the integrable  Toda lattice, \rf{Toda}, or an integrable Birkhoff normal form, \rf{Ford,Rink}, or even some nearby as yet unknown integrable system. Other research directions include the derivation of explicit solutions; in \rf{FrWa}, Friesecke and Wattis proved the existence of solitary waves solutions for a broad range of nonlinearities. This was followed by a series of papers investigating traveling wave solutions of lattices and their approximation by Korteweg--deVries solitons; see \rf{Pankov} for details. See also \rf{ArKoTe} for results on periodic solutions. Another productive line of research has been to investigate what happens when not all the masses and springs are identical, particularly systems with alternating masses and/or spring stiffnesses; see, for example, \rf{BruVer, GGMV, Pankov}.

In an unrelated but also surprising development, in the 1990s, Konstantin Oskolkov, \rf{OskolkovV}, and, independently, Michael Berry  and collaborators, \rf{Berry,BerryKlein,BMS}, discovered the Talbot effect, which the latter named after an 1835 optical experiment, \rf{Talbot}, of the Victorian scientist, inventor, and photography pioneer William Henry Fox Talbot, inventor of the photographic negative.  The Talbot effect arises in quantum mechanics through the behavior of rough solutions to the free space linear Schr\"odinger equation on a circular domain, \ie subject to periodic boundary conditions.  The evolution of a piecewise smooth but discontinuous initial profile, \eg a step function, produces a fractal profile at irrational times (relative to the circumference of the circle) but ``quantizes'' into piecewise smooth but discontinuous profiles at rational times.
Moreover, the fundamental solution, induced by an initial delta function, exhibits ``revivals'' at rational times, localizing into a finite linear combination of delta functions. 
This has the astonishing consequence that, at rational times, the solution to \iz{any} periodic initial value problem is a finite linear combination of translates of the initial data and hence its value at any point on the circle depends only upon finitely many of the initial values!
The effect underlies the experimentally observed phenomenon of quantum revival, \rf{BMS, YeaStr,VVS}, in which an electron, say, that is initially concentrated at a single location of its orbital shell is, at rational times, re-concentrated at a finite number of orbital locations.  

The subsequent rediscovery of this remarkable phenomenon by the first author, in the context of the periodic linearized Korteweg--deVries equation, \rf{Odq, P}, showed that fractalization and quantization phenomena appear in a wide range of linear dispersive \mbox{(integro-)differential} equations, including models arising in fluid mechanics, plasma dynamics, elasticity, DNA dynamics, and elsewhere.  Such linear systems exhibit a fascinating range of as yet poorly  understood dynamical behaviors, whose qualitative features are tied to the large wave number asymptotics of the underlying dispersion relation. 
These studies were then extended, through careful numerical simulations, \rf{COdisp}, to show that fractalization and quantization also appear in a variety of  nonlinear dispersive equations, including integrable models, such as the nonlinear Schr\"odinger, Korteweg--deVries, and modified Korteweg--deVries equations, as well as their non-integrable generalizations with higher degree nonlinearities, \rf{BonaSaut}. (It is fascinating to speculate on possible alternate histories were Zabusky and Kruskal to have conducted their original investigations with discontinuous initial data!)  Some of these numerical observations were subsequently rigorously confirmed in papers of Erdo\utxt gan and collaborators, \rf{ChErTz,ErSh,ErTznls,ErTz}; see also  earlier analytical work of Oskolkov, \rf{OskolkovV}, and Rodnianski, \rf{Rodnianskif}.

Given that the Korteweg--deVries equation and its generalizations arise as continuum models for FPUT chains, the question naturally arises as to whether dispersive fractalization and quantization effects appear in the discrete FPUT system.  Resolving this question is the aim of this paper and its planned sequel(s).  We initially focus our attention on the much simpler linear system, which can be analytically integrated.  We find that, on an appropriately long time scale, the solutions to the periodic linear FPUT chain subject to a step function initial displacement do exhibit a suitably interpreted discrete version of fractalization.   Although FPUT does not exhibit quantization in the sense that KdV does, we do observe coarse-scale similarity between the FPUT and KdV profiles at quantized times, which could be interpreted as ``approximate quantization'', in which the FPUT profile is the superposition of a quantized profile with some sort of fractal ``noise''. On the other hand, we were as yet unable to detect any trace of revival in the discrete  linear system, an observation which remains not entirely understood.  Indeed, there is a noticeable disparity between the FPUT system and its continuum Korteweg--deVries model at such times.  When the system is subjected to a highly concentrated initial displacement --- displacing a single mass in the FPUT system or imposing a delta function in the Korteweg--deVries equation --- at rational long range times the Korteweg--deVries profiles exhibit revival by re-concentrating at a finite number of locations, whereas the linear FPUT profile remains in a similar fractal form as its nearby irrational times. 

In the final section, we describe some numerical investigations in an initial attempt to extend our analysis to fully nonlinear FPUT systems. We briefly survey the use of geometric integrators that have been introduced to numerically integrate FPUT systems, including symplectic integrators such as the St\"ormer/Verlet scheme and trigonometric integrators.  However, we find that these are insufficiently accurate at the high wave numbers that are essential to our investigations, and so turn to a higher-order Hamiltonian splitting scheme to effect the computations.   In this paper, we shall exhibit some preliminary numerical data, for a relatively small number of masses, that indicate these phenomena also appear in nonlinear mass-spring chains when the nonlinearity is sufficiently small. However, owing to the time scales involved and consequent large amount of computation required, we are currently unable to treat chains with a sufficient number of masses that would allow us to definitively address the basic question for nonlinear FPUT systems and generalizations thereof.  And so we defer the further development of more powerful analytical and numerical tools in order to conclusively deal with this intriguing problem.

\Section{FP} Fermi--Pasta--Ulam--Tsingou Chains and their Continuum Models.

The Fermi--Pasta--Ulam--Tsingou (FPUT) system consists of a one-dimensional chain of masses that are connected by springs with nonlinear restoring forces, \rf{FPU,Ford, Pankov}.  We will only consider the case when all the masses and springs are identical.  The dynamics of the mass-spring chain follow immediately from Newton's Laws, taking the form of a system of second order ordinary differential equations for the mass displacements $u_n(t)$ for $n \in \Z$ at time $t$:
\Eq{FPU}
$$\eeq{\om ^{-2}\: \odt{u_n}t = F(u_{n+1} - u_n) - F(u_n - u_{n-1}) \\= u_{n+1} - 2\:u_n + u_{n-1} + N(u_{n+1} - u_n) - N(u_n - u_{n-1}),}$$
where 
$\om $ is the resonant frequency of the linear spring. The forcing function has the form
\Eq F
$$F(y) = y + N(y) = V'(y), \roq{with potential} V(y) = \f2\:y^2 + W(y),$$
where $y$ indicates the elongation of an individual spring.    The nonlinear intermass forcing term is prescribed by $N(y) = W'(y)$. 
As in \rf{Zab,ZabKru}, we will focus attention on the quadratic case when
\Eq{N2}
$$N(y) = \alpha \: y^2,$$
although higher degree polynomials in $y$, particularly cubic, are also of great interest.  Another important system is the integrable Toda lattice, \rf{Toda}, where
\Eq{NToda}
$$V(y) = \alpha \: e^{\beta \: y}.$$
Further examples include the Calogero--Moser integrable system and its trigonometric, hyperbolic, and elliptic generalizations, \rf{Calogero, MoserH, Sutherland}, where
\Eq{CMS}
$$V(y) = \frac\alpha{y^2} \orx V(y) = \frac\alpha{\sin^2 y} \orx V(y) = \frac\alpha{\sinh^2 y} \orx V(y) = \CP(y),$$
with $\CP$ the Weierstrass elliptic function, as well as the Lennard--Jones potential, \rf{LJ}, 
\Eq{LJ}
$$V(y) = \frac\alpha{y^{12}} - \frac\beta{y^{6}} ,$$
which is used to model interatomic and intermolecular dynamics.
 
In this note, we will concentrate on the periodic problem,  viewing the system as a circular chain consisting of $\M$ masses, which are labelled so that $u_{n_1}(t) = u_{n_2}(t)$ whenever $n_1 \equiv n_2 \>\mod \M$.   Alternatively, the case of an infinite chain, where the displacements of the masses suitably decay at large distances, is important.  However, since the continuum models have smooth evolutionary behavior on the line --- dispersive quantization being intimately tied to the periodic boundary conditions --- we do not anticipate any unexpected effects in an infinite FPUT chain and so do not pursue it here.  The \is{Dirichlet problem}, in which the first and last masses are pinned down, so that $u_0(t) = u_\M(t) = 0$, with  $\rgo n \M$, is also of interest, \rf{Zab}, but its analysis will be deferred to subsequent investigations.

Following \rf{ZabKru,Zab,RFPU, RFPUB}, we endeavor to better understand the discrete FPUT dynamics by passing to a continuum model.  To this end, we assume the masses lie on a circle of fixed radius, say the unit circle of circumference $2\pii$.  As the number of masses $\M \to \infty $, the equilibrium intermass spacing $h = 2\pii/\M \to 0$.  To maintain consistency, the time must be correspondingly rescaled, $t \mapsto h\:t$, and so we consider the system
\Eq{rFPU}
$$ \odt{u_n}t = \frac{\cm^2}{h^2} \bbk{F(u_{n+1} - u_n) - F(u_n - u_{n-1})},$$
where $\cm = \om \:h$ will be the wave speed of the limiting scalar wave equation; see below.

We can view the individual displacements as the sample values of an interpolating  function $u(t,x)$ that is $2\pii$ periodic in $x$, 
so that 
$$u_n(t) = u(t,x_n), \where x_n = n\:h = \frac{2\pii\:n}\M, \qquad n \in \Z,$$
are the \emph{nodes} or reference positions of the masses.
To produce a continuum model, we apply Taylor's theorem to expand
$$u_{n\pm 1}(t) = u(t, x_n \pm h) = u \pm h\: u_x + \f2\:h^2 u_{xx} \pm \f6\:h^3 u_{xxx} + \cdotsx,$$
where the right hand side is evaluated at $(t,x_n)$.  Substituting into \eq{FPU} and replacing $x_n \mapsto x$, we arrive at the dispersive partial differential equation
\Eq{dpde}
$$u_{tt} = \cm^2\paz{K\br u + Q \br u},$$
with linear component 
\Eq K
$$K\br u = u_{xx} + \f{12} h^2 u_{xxxx} + \Oh4,$$
while $Q\br u$ is obtained by similarly expanding the nonlinear terms.  For example, in the quadratic case\fnote{For unexplained reasons, Zabusky, \rf{Zab}, is missing the term involving $\dsty u^{\,}_{x} u^{\,}_{xxxx}$.}
\eq{N2},
\Eq Q
$$Q \br u = 2\: \alpha \: h\: u_x u_{xx} + \f6\: \alpha \: h^3 \paz{u_{x} u_{xxxx} + 2\: u_{xx} u_{xxx}} + \Oh5.$$
Thus, assuming the linear wave speed $\cm$ and nonlinear scale parameter $\alpha$ are both $\O1$, we obtain, to second order in $h$, the bidirectional continuum model
\Eq{FPUc2}
$$u_{tt} = \cm^2 \paz{u_{xx} + 2\: \alpha \: h\: u_x u_{xx} + \f{12} h^2 u_{xxxx}},$$
a potential form of the integrable (nonlinear) \is{Boussinesq equation}, \rf{DJ}, 
\Eq{iBeq}
$$v_{tt} = \cm^2 \paz{v_{xx} + \alpha \: h\: (v^2)_{xx} + \f{12} h^2 v_{xxxx}},$$
which can be obtained by differentiating \eq{FPUc2} with respect to $x$ and replacing $u_x\mapsto v$.
Note that, to leading order, the continuum model \eq{FPUc2} coincides with the standard linear wave equation $u_{tt} = \cm^2 u_{xx} $ with wave speed $\cm>0$.  
The \is{Korteweg--deVries equation} is obtained through a standard ``unidirectional factorization'' of the preceding bidirectional system, \rf{Whitham}, \ie assuming the waves are only propagating in one direction, say in the direction of increasing $x$, producing
\Eq{kdv}
$$u_t + \cm\,\paz{u_x +  \alpha \: h\: u\: u_x + \f{24} h^2 u_{xxx}} = 0.$$
In more detail, one begins with the d'Alembert formula, \rf P, that represents the solutions to the linear wave equation $u_{tt} = c^2 u_{xx} $ as a linear combination of of left and right-moving waves, that, respectively solve the first order ``factors'' $\myeq{u_t = c u_x, \\ u_t = -\:c\: u_x}$.  For a higher order or nonlinear perturbation, one seeks an expansion of, say, the right moving factor in powers of the small parameters, where the individual terms, such as those appearing in the Korteweg--deVries equation \eq{kdv}, are uniquely determined by requiring the expansion be consistent with the underlying bidirectional model. See \rf{Whitham} for details.

\medskip

\Rmk For the cubically forced FPUT system, the unidirectional model is the integrable modified Korteweg--deVries equation, \rf{RFPUB,Zab}, in which the nonlinear term is a multiple of $u^2 u_x$.  Higher degree polynomial forcing functions produce generalized Korteweg--deVries equations with higher degree nonlinearities, which are no longer integrable, and, in fact, can induce blow up of solutions, \rf{BonaSaut}.
 
To initiate our investigations, let us ignore the nonlinear contributions and concentrate on the linear FPUT system and its  continuum models.   The rescaled linear system becomes what is known as the \is{discrete wave equation}, \rf{Pankov}:
\Eq{LFPU}
$$\odt{u_n}t = \frac{\cm^2}{h^2} \paz{u_{n+1} - 2\:u_n + u_{n-1}}.$$
 Note that the parameter $\cm^2/h^2$ can be set to unity by further rescaling time, but for comparative purposes with both the linearized continuum models and their nonlinear counterparts, it is important to leave it in. 
The discrete wave equation, to the same order in $h$, has bidirectional continuum model
\Eq{bBe}
$$u_{tt} = \cm^2 \paz{u_{xx} + \f{12}\: h^2 u_{xxxx}},$$
known as the linearized ``bad Boussinesq equation'', \rf{RFPU}, owing to the fact that it is an ill-posed partial differential equation.
Indeed, its dispersion relation is found by the usual method, \rf{Whitham}, of substituting the exponential  ansatz
\Eq{exp}
$$u(t,x) = e^{\i (k\:x - \omega \:t)},$$
producing the algebraic equation
\Eq{bBed}
$$ \omega ^2 = p_4(k) = \cm^2 k^2\paz{1 - \f{12} \:h^2 k^2}$$
relating the temporal frequency $\omega $ to the wave number (spatial frequency) $k$.
Because $p_4(k) < 0$ for $k \gg 0$, the bad Boussinesq model \eq{bBe} is \iz{not} purely dispersive since the high wave number modes induce complex conjugate purely imaginary values of $\omega $ and hence exponentially growing modes, \rf{DarHua}, that underly the ill-posedness of the initial value problem.  
Interestingly, the corresponding linearized unidirectional Korteweg--deVries model 
\Eq{lkdvh}
$$u_t + \cm\,\paz{u_x +  \f{24} h^2 u_{xxx}} = 0$$
does not suffer from this instability, since its dispersion relation
\Eq{lkdvhd}
$$\omega = \cm \:k\paz{1 - \f{24} \:h^2 k^2},$$
 is everywhere real, and coincides with the Taylor expansion, at $k=0$, of one of the two branches of the bidirectional dispersion relation  \eq{bBed}. 
 
There are three common mechanisms for overcoming the illposedness of the Boussinesq model, leading to slightly different well-posed models that all agree to order $h^2$.  For completeness, we present these models next, noting that the second and third regularized models exhibit very similar behavior in the present context. 

 The first way of regularizing the bad Boussinesq model is to replace it by the linearized sixth order \is{bidirectional Korteweg--deVries equation}
\Eq{bKdV}
$$u_{tt} = \paz{\cm \:\partial _x + \f{24} \:\cm \:h^2 \partial _x^3}^2 u = \cm^2 \paz{u_{xx} + \f{12}\: h^2 u_{xxxx} + \f{576}\: h^4 u_{xxxxxx}},$$
which agrees to order $h^2$.  It has solutions that are an exact linear combination of right- and left-moving linear KdV solutions.  Positivity of the right hand side of the corresponding dispersion relation
\Eq{bKdVd}
$$ \omega ^2= \cm^2 k^2\paz{1 - \f{24} \:h^2 k^2}^2 =  \cm^2 k^2\paz{1 - \f{12} \:h^2 k^2 + \f{576}\: \:h^4 k^4}$$
implies well-posedness of this sixth order model.

\medskip

\Rmk It remains somewhat mysterious to the authors how an ill-posed bidirectional wave model can have right- and left-moving unidirectional constituents that are both well-posed.  This is clearly a consequence of the use of low wave number Taylor expansions of the dispersion relation near $k=0$ in the approximation procedure, but we would argue that this seeming paradox warrants further study.  

Another means of regularizing the linear, and hence the nonlinear, continuum model is to retain the order $h^4$ terms in the preceding Taylor expansion.  This produces the sixth order linear partial differential equation 
\Eq{LFPUc6}
$$u_{tt} = \cm^2 \paz{u_{xx} + \f{12}\: h^2 u_{xxxx} + \f{360}\: h^4 u_{xxxxxx}},$$
with dispersion relation
\Eq{FPUd6}
$$ \omega ^2 = p_6(k) = \cm^2 k^2\paz{1 - \f{12} \:h^2 k^2 + \f{360}\: \:h^4 k^4}.$$
Since $p_6(k) > 0$ for all $k$, the regularized model \eq{LFPUc6} is purely dispersive, and hence well-posed, in that all Fourier modes maintain their form under translation.

An alternative regularization procedure that avoids increasing the order of the differential equation is to replace two of the $x$ derivatives in the fourth order term in the bad Boussinesq equation \eq{bBe} by $t$ derivatives\fnote{A similar device is used to derive the BBM or Regularized Long Wave model, \rf{BBM,Whitham}, which is a non-integrable alternative to the Korteweg--deVries equation that nevertheless has nicer functional analytic properties.  See also \rf{OHamwwa, OHamwwb} for a variety of related higher order models for shallow water waves.}, using the fact that, to leading order, $u_{xx} = \cm^{-2} u_{tt} + \Oh2$, thereby producing the continuum model
\Eq{rBe}
$$u_{tt} = \cm^2 u_{xx} + \f{12}\: h^2 u_{xxtt},$$
known as the \is{linear Boussinesq equation}, \rf{Whitham; {pp. 9, 462}}.  It arises as the linearization of Boussinesq's bidirectional model for shallow water waves, and has also been proposed as a model for DNA dynamics, \rf{Scott}.
In this case, the dispersion relation is
\Eq{rBed}
$$ \omega ^2 = q(k) = \frac{\cm^2 k^2}{1 + \f{12} \:h^2 k^2}\approx \cm^2 k^2\paz{1 - \f{12} \:h^2 k^2 + \f{144}\: \:h^4 k^4 + \cdotsx} > 0,$$
and hence the equation is purely dispersive and well-posed.

Following this presentation of the continuum models, let us now derive the analogous ``dispersion relation'' for the discrete linearized FPUT system (discrete wave equation) \eq{LFPU}; see also \rf{ZabDeem}. Substituting the usual exponential ansatz \eq{exp}, evaluated at the node $x = x_n = n\:h$, into \eq{LFPU} produces
\Eq{dFPU}
$$\eeq{ -\:\omega ^2 e^{\i (k\:x_n - \omega \:t)}= \frac{\cm^2}{h^2} \paz{e^{\i (k\:x_n + k\:h- \omega \:t)} -2\: e^{\i (k\:x_n - \omega \:t)} + e^{\i (k\:x_n - k\:h - \omega \:t)}}\\
= -\,\frac{2\:\cm^2}{h^2}\pa{1 - \cos k\:h}\,e^{\i (k\:x_n - \omega \:t)} .}$$
We thus deduce the discrete FPUT dispersion relation
\Eq{FPUd}
$$\eeq{\omega ^2  = \frac{2\:\cm^2}{h^2} \pa{1 - \cos k\:h}
= \frac{4\:\cm^2}{h^2} \sin^2 \f2\:k\:h = \frac{\cm^2\M^2}{\pi^2} \sin^2 \frac{k\pii}\M}$$
that determines the temporal frequencies $\omega $ in terms of the wave numbers $k$.
Since $\omega (k)$ is real for all $\rgo k\M$, the FPUT system can be regarded as dispersive, in that the different Fourier modes propagate unchanged at different wave speeds.  This implies ``well-posedness'' or, more accurately, since we are dealing with a system of ordinary differential equations, stability of the equilibrium solution.
Moreover, observe that the continuum model dispersion relations \eqasss{bBed}{FPUd6}{bKdVd}{rBed} all have the same order $h^4$ Taylor expansion at $k=0$ as \eq{FPUd}, and hence approximate it well at low or even moderately large wave numbers.  However, they exhibit rather different high wave number asymptotics, which, as noted in \rf{COdisp}, is the key property that governs the dispersive fractalization of rough solutions.

\Section R The Riemann Problem.   

As in \rf{COdisp,Odq}, we are particularly interested in the \is{Riemann problem} --- an initial value problem of fundamental importance in the study of hyperbolic wave equations and shock waves, \rf{Smoller}.
Here, the initial displacement is a (periodically extended) step function:
\Eq{step}
$$u(0,x) = \tcases{1,& 0 < x < \pi,\\0,& -\pii < x < 0,\\\f2,& x = -\pii,\> 0,\> \pi,}$$
whose values at integer multiples of $\pi$ are specified in accordance with the convergence properties of its Fourier series
\Eq{Fstep}
$${u(0,x) = \fra2 + \frac 2\pi\; \sum _{ \text{odd } k = 1 } ^\infty \frac{\sin {k\:x}}{k}\,.}$$
We will also, for simplicity, impose zero initial velocity, concentrating on the pure displacement problem: 
\Eq{v0}
$$u_t(0,x) = 0,$$
leaving the analysis of the effects of a nonzero initial velocity to a subsequent study.
Given a well-posed linear bidirectional continuum model equation, such as \eqs{bKdV}{LFPUc6}, or \eq{rBe}, with associated dispersion relation $\omega (k)$, the Fourier series for the  solution to the corresponding periodic initial value problem \eqc{step}{v0} takes the form of a linear combination of standing wave solutions:
\Eq{stepsol}
$${u(t,x) = \fra2 + \frac 2\pi\; \sum _{ \text{odd } k = 1 } ^\infty \frac{\cos \omega (k)\,t \ \sin {k\:x}}{k}\,.}$$
By an elementary trigonometric identity, we can split the standing wave summands in \eq{stepsol} into right- and left-moving unidirectionally propagating waves:
\Eq{RL}
$$u(t,x) = \frac{u_R(t,x) + u_L(t,x)\sstrut5}2,$$
where the factor of $\f2$ is introduced for comparative purposes, ensuring that  all three solutions have the same initial displacement:
$$u(0,x) = u_R(0,x) = u_L(0,x).$$
The right-moving constituent is
\Eq{stepsolR}
$${u_R(t,x) = \fra2 + \frac2\pi\; \sum _{ \text{odd } k = 1 } ^\infty \frac{\sin \bpa{k\:x - \omega (k)\,t}}{k}\,,}$$
and its left-moving counterpart is obtained by replacing $t$ by $-\:t$.  A key feature of such series solutions that produces the dispersive fractalization and quantization effects is the slow decay of their Fourier coefficients, which implies that they are conditionally but not absolutely convergent.

The corresponding step function initial data for the discrete FPUT problem is obtained by sampling \eq{step} at the nodes.  To make the connection, we will take the number of masses to be even, $\M = 2\:\m$, and the nodes to be 
$$\qeq{x=x_n = n\:h = \pi\:n/\m, \\\rgs n{-\:\m}m,}$$
identifying $x_{-\m} = x_m$ by periodicity.  Thus, the initial data for the Riemann problem for the FPUT system is given by
\Eq{dstep}
$$u_n(0) = \tcases{1,& 0 < n < \m,\\0,& -\:\m < n < 0,\\\f2,& n = -\:\m,\ 0,\ \m.}$$
In other words, we displace each mass lying on the ``right semicircle''  by $1$ unit, while those on the left remain at their equilibrium position, except the two masses lying at the interface that are displaced by only half a unit.  
As in \eq{v0}, the masses are assumed to be at rest initially:
\Eq{ud0}
$$\udot_n(0) = 0.$$
We use the Discrete Fourier Transform to write the solution as a 
Fourier sum
\Eq{DFT}
$$u(t,x) \sim \Summ k{1-m}m c_k(t) e^{\i k\:x},$$
over the fundamental periodic modes, \rf{OS; Section 5.6}, where the symbol $\sim$ will mean that the left and right hand sides agree at the nodes, \ie when $x= x_n$.  
The Discrete Fourier Transform applied to the sampled step function produces the interpolating discrete Fourier sum
\Eq{DFTstep}
$$u(0,x) \simx \fra2 + \fra \m\; \sum _{ \text{odd } k = 1 } ^m  \cot \f2\:k\:h\ \sin k\:x.$$
In view of the dispersion relation \eq{FPUd}, the resulting solution to the linearized FPUT chain is
\Eq{FPUss}
$$\aeq{u(t,x) &\simx \fra2 + \fra \m\; \sum _{ \text{odd } k = 1 } ^m  \cot \f2\:k\:h\ \cos \omega (k)\,t \ \sin k\:x\\
&\eqx \fra2 + \fra \m\; \sum _{ \text{odd } k = 1 } ^m  \cot \f2\:k\:h\ \cos \Pa{ \frac{ 2\:c\:t }{ h } \sin \f2\:k\:h} \sin k\:x,}$$
meaning that the displacement of the \nth mass is given by sampling the right hand side at the nodes: 
\Eq{DFTsample}
$$\req{u_n(t) = u(t,x_n),\\x_n = n\: h = \frac{\pi\:n}\m\,.}$$
Again, the solution \eq{FPUss} is a linear combination of standing waves, and can be decomposed into left and right moving constituents, as in \eq{RL}.
The right-moving constituent has the explicit form
\Eq{FPUssR}
$$u_R(t,x) \simx \fra2 + \fra \m\; \sum _{ \text{odd } k = 1 } ^m  \cot \f2\:k\:h\ \sin \Pa{ k\:x - \frac{ 2\:c\:t }{ h } \sin \f2\:k\:h}.$$
As above, its left-moving counterpart is obtained by replacing $t$ by $-\:t$.

\medskip

\Rmk Note that if we omit the constant term, the bidirectional solutions constructed in \eqc{stepsol}{FPUss} also satisfy Dirichlet  boundary conditions, with a half-size signum function as initial condition:  $u(t,x) = \f2\:\sgn x\ostrut{9}6$ for $ -\pii < x < \pi$.  Thus, all our subsequent remarks on their behavior also apply to this Dirichlet initial-boundary value problem. On the other hand, their unidirectional constituents do not individually satisfy Dirichlet boundary conditions.

As shown in \rf{Odq,OskolkovV}, the canonical linearized Korteweg--deVries equation 
\Eq{ckdv}
$$u_\tau + u_{\xi \xi \xi } = 0$$
with step function initial data and periodic boundary conditions on $-\pii \leq \xi \leq \pi$ exhibits dispersive fractalization and quantization in the following sense.  At \is{irrational times} $\tau > 0$, meaning $\tau/\pi \not \in \Q$, the solution profile $u(\tau,\xi)$  is a continuous but non-differentiable fractal.  On the other hand, at rational times, $\tau/\pi \in \Q$, the solution is discontinuous, but piecewise constant!  Indeed, if $\tau = 2\pii\:p/q$ where $p,q \in \Z$ have no common factors, then the solution is constant on the intervals $2\pii j/q < \xi < 2\pii (j+1)/q$ for $j \in \Z$.  Thus, the larger the denominator $q$, the shorter the intervals of constancy.  (It is possible that the solution achieves the same constant value on one or more adjacent intervals, and so an interval of constancy may be larger than specified above.  See \rf{OTpcon} for a number-theoretic investigation into when this occurs.)  A rigorous proof of the fractal nature of the solution at irrational times, including the estimate that its fractal dimension $d$ is bounded by $\fr32 \leq d \leq \fr74\sstrut5$, can be found in \rf{ErSh}.

Indeed, the results of \rf{ErSh} imply that a linear evolutionary integro-differential equation with dispersion relation that is (in an appropriate sense) asymptotic to a power of $k$ at large wave numbers, $\omega (k) \sim k^\alpha $ as $\ki$, for $1 \ne \alpha > 0$, exhibits fractalization at almost all times, provided the initial data is of bounded variation, but not too smooth, meaning it does not lie in any Sobolev space $H^{\beta }$ for $\beta > \f2\sstrut5$, \ie its Fourier coefficients $c_n$ decay sufficiently slowly so that the series $\sum\; (1+n^2)^{\beta } \>\abs{c_n}^2\sstrut5$ diverges.
Moreover, if the asymptotic exponent $\alpha $ is integral, $2 \leq \alpha \in \Z$,  numerical experiments, \rf{COdisp}, indicate that the solution profiles quantize at other times, in the sense that they take a different form from the ``generic'' fractal profiles: piecewise smooth with either jumps or cusps, possibly with some much smaller fractal modulation superimposed.  However, being so far based on numerical calculations, it is not yet known if the observed small scale fractals on the quantized profiles are genuine or just a manifestation of numerical error.  See the recent preprint \rf{BOPS} for further interesting developments.

\Warning The fractal dimension of the graph of a function can be misleading.  For example, the graph of the sinusoidal function $f(x) = \sin(1/x)$ has fractal dimension $2$ even though it is perfectly smooth, even analytic, except at the singularity at $x=0$.  For this reason, the results in \rf{ErSh}, while striking, are, on a deeper level, unsatisfying.  It would be preferable to know, or at least have estimates on, the Hausdorff dimension of (sections of) such solution profiles; however this seems to be beyond current analytic capabilities.

Turning our attention to the linearized Korteweg--deVries model \eq{lkdvh}, the leading first order term $c\:u_x$ represents linear transport moving at speed $-\:c$, and only affects each solution profile by an overall translation.  We can map \eq{lkdvh} to the preceding canonical form
\eq{ckdv} by a Galilean shift to a moving coordinate frame, coupled with a rescaling of the time variable: 
\Eq{xitau}
$$\req{ \xi = x - c\:t,\\ \tau = \f{24}\:c\:h^2\:t.}$$
Due to the temporal scaling, the dispersive quantization pattern occurring in the solution to the canonical KdV \eqe{ckdv} at a rational time $\tau = 2\pii\:p/q$ will now appear (suitably translated) at a much later time, namely $t = 48\pii \: p/(c\:h^2 q)$.  Since the normalized model \eq{ckdv} exhibits quantization at \iz{every} rational time, the same is true (modulo the scaling factor used to distinguish rational from irrational) of the FPUT model version \eq{lkdvh}.  However, in the latter model, at a rational time $t = \O1$, the denominator $q$ will be very large, of order $\Oh{-2}$, hence the intervals of constancy are extremely small, $\Oh2$, and thus undetectable at the physical level, which has spatial scale  $\Delta x = \O h$.  At such scales, both physical and graphical, it will be practically impossible to distinguish such profiles from  fractals. 

\begin{figure}
  \hfil
  \begin{minipage}[t]{1.75in}
    \centering
    \includegraphics[width=\textwidth]{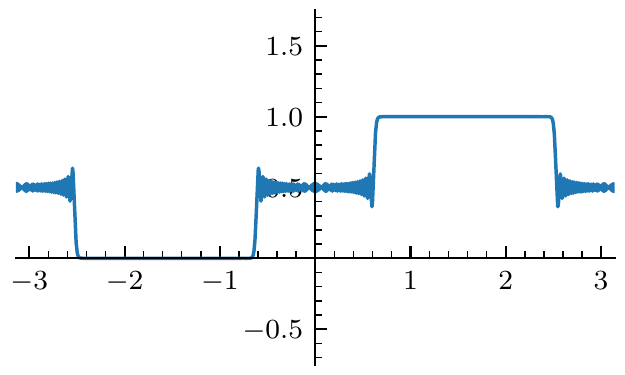}
    \includegraphics[width=\textwidth]{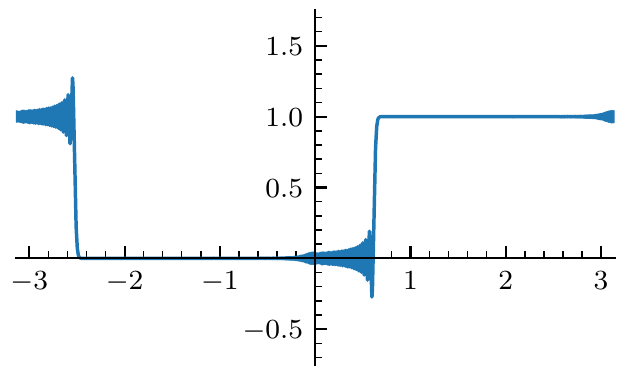}
    KdV
  \end{minipage}
  \hfil
  \begin{minipage}[t]{1.75in}
    \centering
    \includegraphics[width=\textwidth]{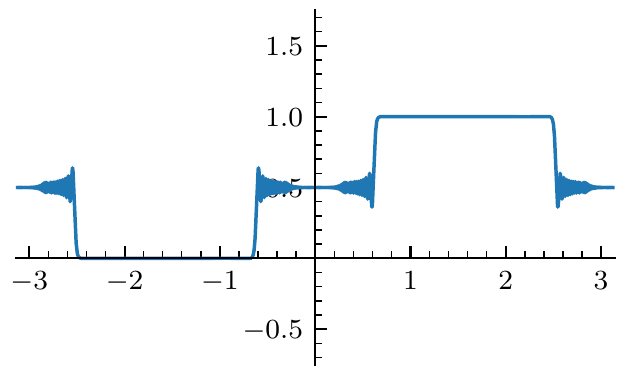}
    \includegraphics[width=\textwidth]{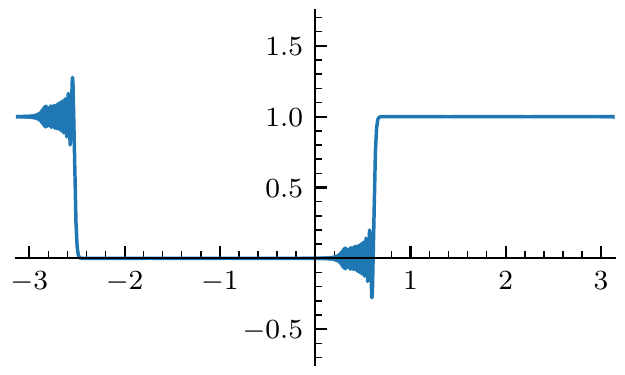}
    Boussinesq
  \end{minipage}
  \hfil
  \begin{minipage}[t]{1.75in}
    \centering
    \includegraphics[width=\textwidth]{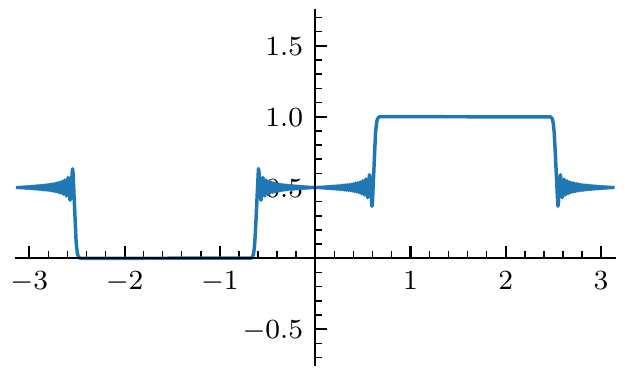}
    \includegraphics[width=\textwidth]{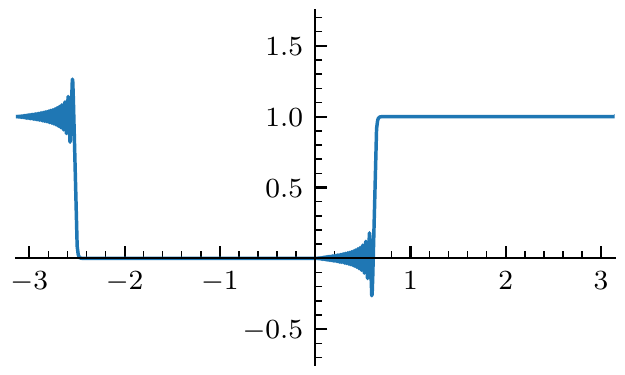}
    FPUT
  \end{minipage}
  \hfil
  \caption{Bi- and uni-directional solution profiles at
    $t=\f5\:\pi$.\label{bup5}}
\end{figure}

Now let us compare the solution to the periodic Riemann initial value problem for the discrete linear Fermi--Pasta--Ulam--Tsingou system \eq{LFPU} with those of the three well-posed linear model equations: the bidirectional Korteweg--deVries model \eq{bKdV}, the sixth order model \eq{LFPUc6}, and the regularized Boussinesq model \eq{rBe}.  For our numerical comparisons, we sum the same modes in the discrete and continuous Fourier series, truncating at $k=m$, and plot the resulting profile; in other words, we are performing exact (well, modulo floating point round off)  computations on the truncated (discrete) Fourier series and not a numerical approximation.  The initial data is a (periodically extended) step function.  In the continuum models, one can work with either the continuous Fourier series representation of the initial data \eq{Fstep}, or the corresponding discrete Fourier sum \eq{DFTstep}.  However, in the given situations, we observe no appreciable differences between the associated solution profiles, and hence will use the discrete version in all figures representing solutions to the Riemann initial value problem.  We fix the wave speed $c=1$, and, for most of our numerical investigations, work with $\m = 512$, so there are $\M = 1024$ masses and $h = \pi/\m \approx .006136$.  Solutions for other numbers of masses have been calculated, and the overall conclusions are similar, although the fewer their number, the less pronounced some effects tend to be.

The first observation is that, on the time scale and resolution under consideration, there is almost no noticeable difference between the sixth order and regularized Boussinesq models, and so we choose only to display the results for the latter.  We will plot both the bidirectional solution $u(t,x)$, as given in \eqs{stepsol}{FPUss} and its unidirectional right-moving constituent \eqs{stepsolR}{FPUssR}.  We consider the effects at times that are selected from three regimes: what we will call \is{short times}, where $t = \O1$, \is{medium times}, where $t = \Oh{-1}$, and \is{long times}, where $t = \Oh{-2}$. 

\begin{figure}
  \hfil
  \begin{minipage}[t]{1.75in}
    \centering
    \includegraphics[width=\textwidth]{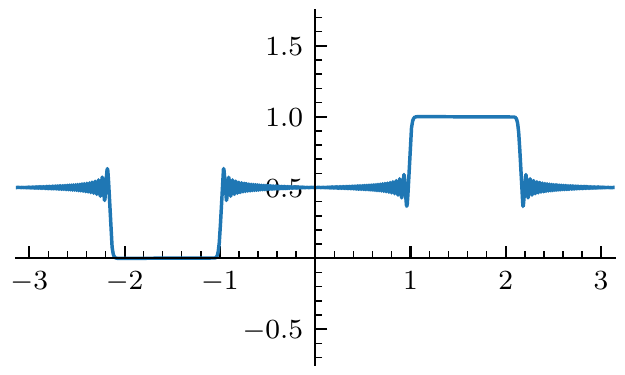}
    \includegraphics[width=\textwidth]{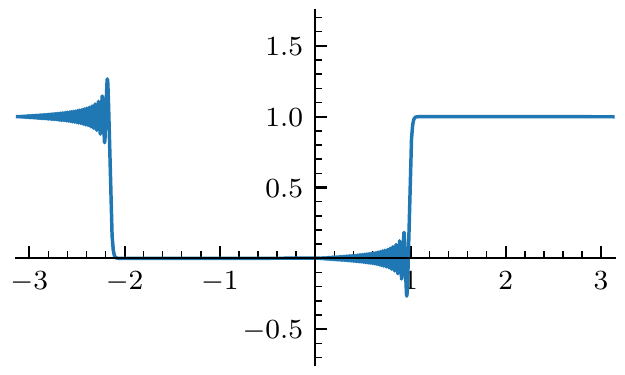}
    $ t = 1 $ 
  \end{minipage}
  \hfil
  \begin{minipage}[t]{1.75in}
    \centering
    \includegraphics[width=\textwidth]{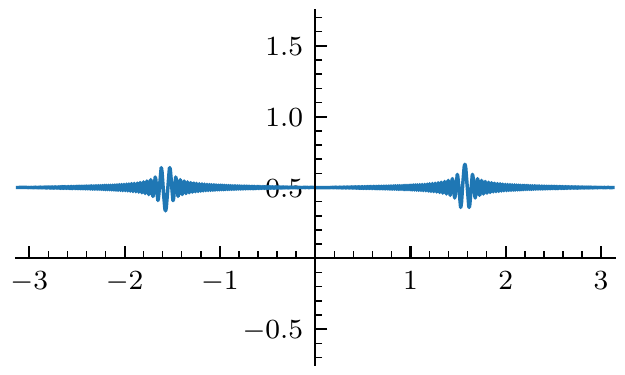}
    \includegraphics[width=\textwidth]{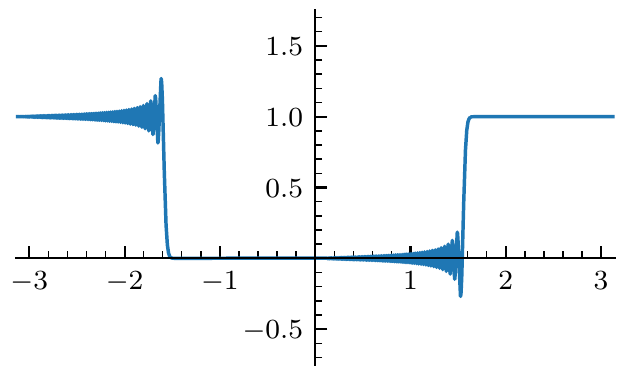}
    $ t = \f2\pii $
  \end{minipage}
  \hfil
  \begin{minipage}[t]{1.75in}
    \centering
    \includegraphics[width=\textwidth]{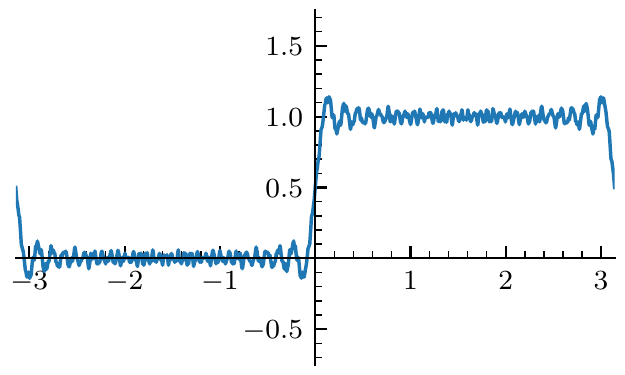}
    \includegraphics[width=\textwidth]{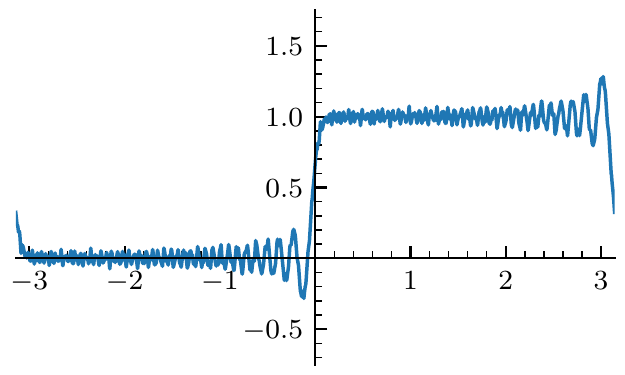}
    $ t = 12\pii $
  \end{minipage}
  \hfil
  \caption{Bi- and uni-directional FPUT solution profiles at short
    times.\label{bu1}}
\end{figure}

First, on short time scales, the solutions to all four models exhibit  little appreciable difference.  For example, consider the profiles at $t=\f5\:\pi$ graphed in \fg{bup5} --- the top row being the full bidirectional solution and the bottom row its right-moving unidirectional constituent.  
Since all profiles remain rather similar at short times, in \fg{bu1} we just graph the FPUT solution profiles. 
What we observe is that, on the short time scale, the solution is an oscillatory perturbation of the traveling wave solution to the corresponding limiting bi- and uni-directional wave equations\fnote{Recall that we have set $c = 1$.} 
\Eq{w2}
$$\req{u_{tt} = u_{xx},\\u_t + u_x = 0.}$$  
In particular, at $t=\f2\pii$, the right- and left-moving waves have cancelled each other out, leaving only a constant solution profile for the traveling wave solution, with a superimposed fractal residue in the FPUT system and its continuum models, all three of which take on a comparable form.

\begin{figure}
  \hfil
  \begin{minipage}[t]{1.75in}
    \centering
    \includegraphics[width=\textwidth]{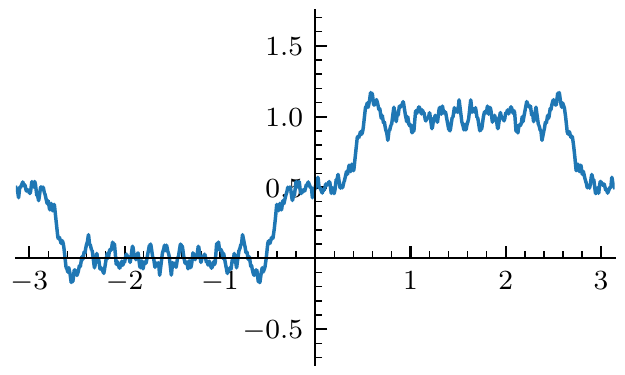}
    \includegraphics[width=\textwidth]{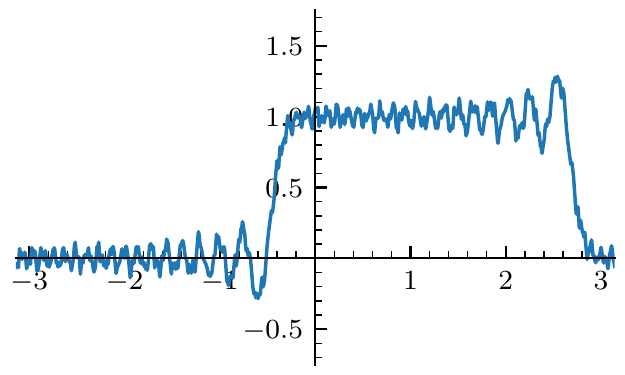}
    $ t = 1/h $ 
  \end{minipage}
  \hfil
  \begin{minipage}[t]{1.75in}
    \centering
    \includegraphics[width=\textwidth]{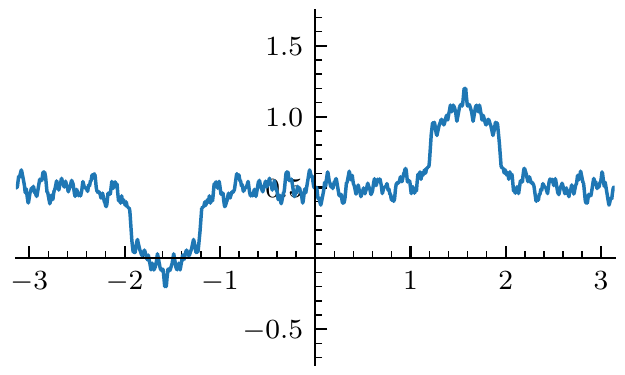}
    \includegraphics[width=\textwidth]{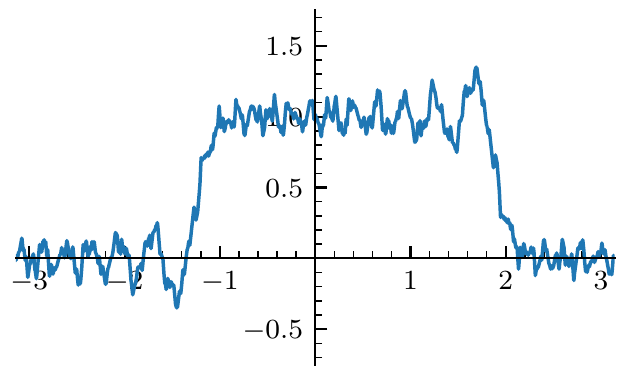}
    $ t = 3/h$
  \end{minipage}
  \hfil
  \begin{minipage}[t]{1.75in}
    \centering
    \includegraphics[width=\textwidth]{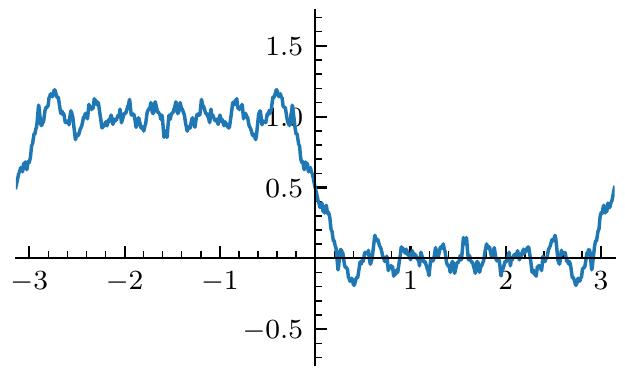}
    \includegraphics[width=\textwidth]{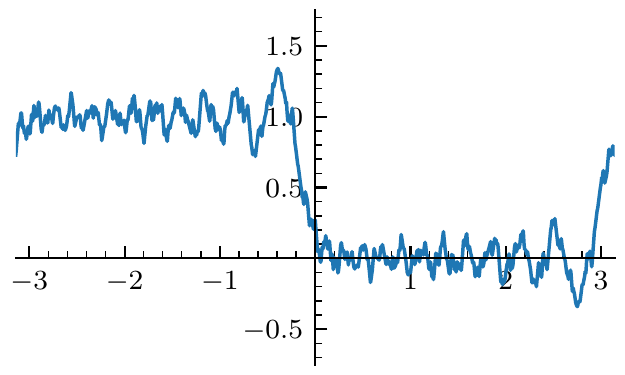}
    $ t = \pi/h $
  \end{minipage}
  \hfil
  \caption{Bi- and uni-directional FPUT solution profiles at medium
    times.\label{buh}}
\end{figure}

At medium times, of order $\Oh{-1}$, the fractal nature of the oscillations superimposed upon the traveling wave solution profile has become more pronounced.  Again, both the FPUT system and its continuum models exhibit  similar behavior; \fg{buh} graphs the former at some representative medium times.  When we decrease the number of masses, the unidirectional profiles look fairly similar  modulo translation due to a change in the average wave speed. The bidirectional profiles look different, because translating the unidirectional profiles leads to different interference patterns. The overall amplitude of the superimposed fractal oscillations remains similar, but their frequencies are related to the number of masses; in other words, when more masses are present, more high frequency modes are excited.

\begin{figure}
  \hfil
  \begin{minipage}[t]{1.75in}
    \centering
    \includegraphics[width=\textwidth]{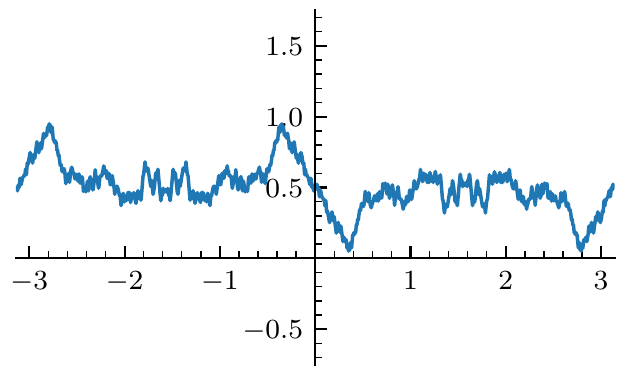}
    \includegraphics[width=\textwidth]{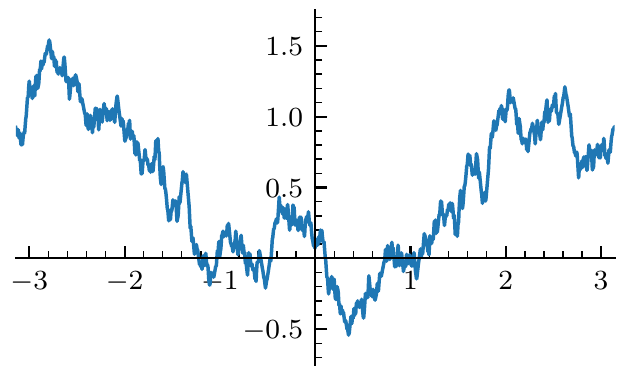}
    KdV
  \end{minipage}
  \hfil
  \begin{minipage}[t]{1.75in}
    \centering
    \includegraphics[width=\textwidth]{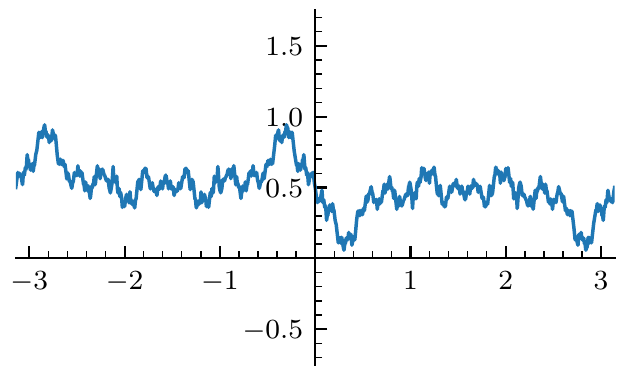}
    \includegraphics[width=\textwidth]{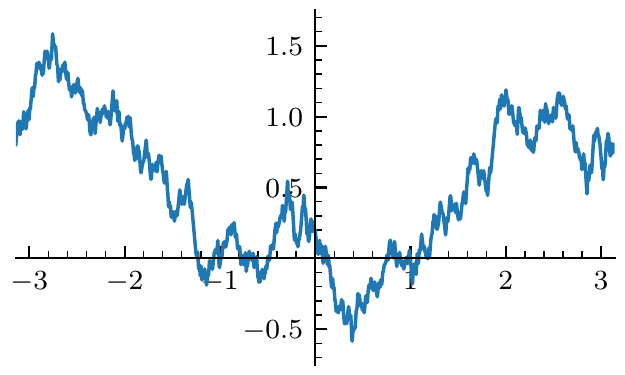}
    Boussinesq
  \end{minipage}
  \hfil
  \begin{minipage}[t]{1.75in}
    \centering
    \includegraphics[width=\textwidth]{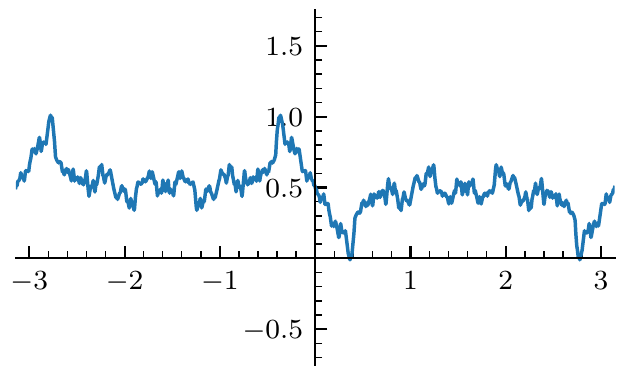}
    \includegraphics[width=\textwidth]{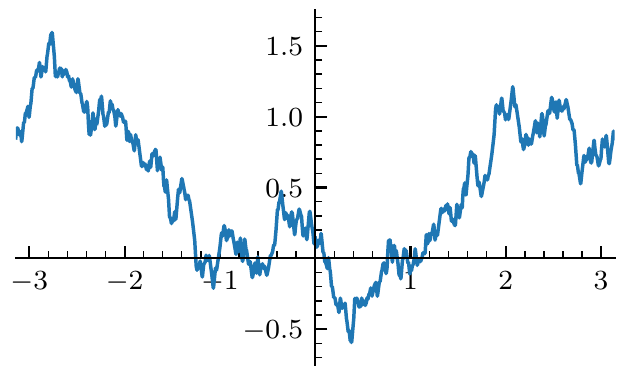}
    FPUT
  \end{minipage}
  \hfil
  \caption{Bi- and uni-directional solution profiles at
    $t=1/h^2\approx 26,\!561$.\label{buh2}}
\end{figure}

Once we transition to the long time scale, of order $\Oh{-2}$, significant differences arise in the observed behaviors.  First let us consider the solution profiles at the irrational (meaning that $h^2\: t/\pi \not \in \Q$) times $t = 1/h^2  \approx 26561$ and $t = 400000$, plotted in \fgas{buh2}{bu3}. All three profiles are of a similar fractal form, albeit with differences in their small scale features.  The unidirectional constituents are more uniformly fractal, while the bidirectional solutions exhibit some semi-coherent regions, perhaps indicating some remnant of the intervals of constancy of a nearby rational profile. 

\begin{figure}
  \hfil
  \begin{minipage}[t]{1.75in}
    \centering
    \includegraphics[width=\textwidth]{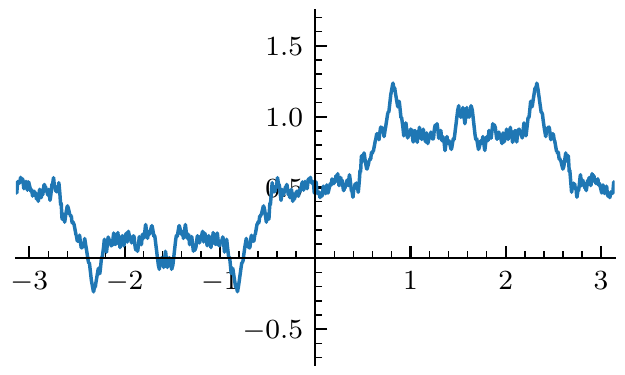}
    \includegraphics[width=\textwidth]{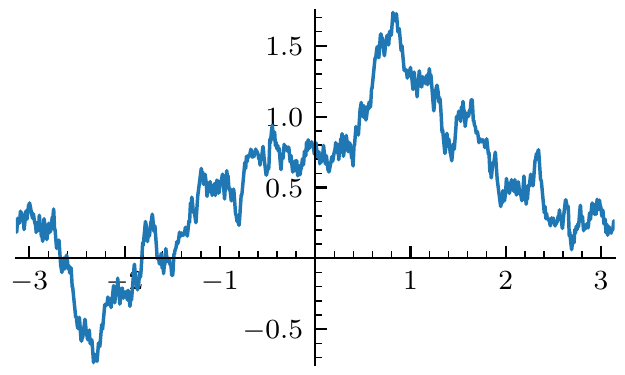}
    KdV
  \end{minipage}
  \hfil
  \begin{minipage}[t]{1.75in}
    \centering
    \includegraphics[width=\textwidth]{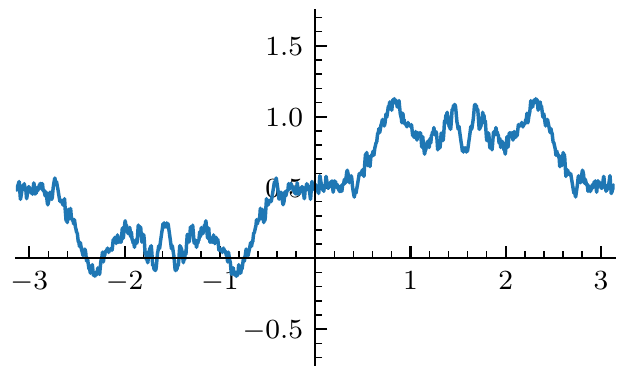}
    \includegraphics[width=\textwidth]{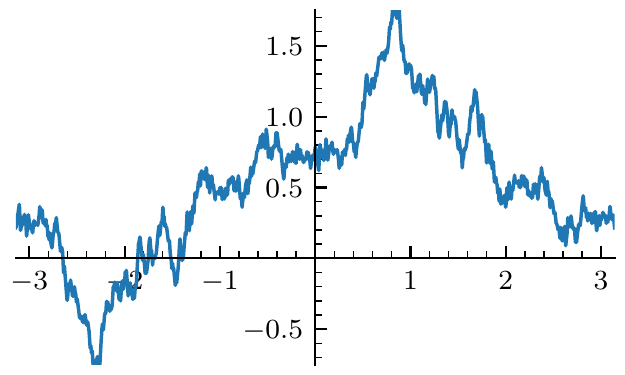}
    Boussinesq
  \end{minipage}
  \hfil
  \begin{minipage}[t]{1.75in}
    \centering
    \includegraphics[width=\textwidth]{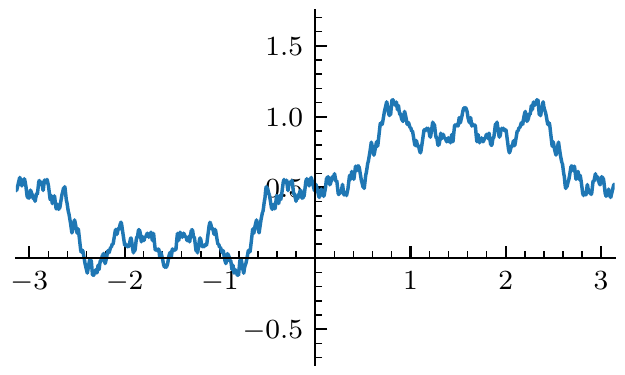}
    \includegraphics[width=\textwidth]{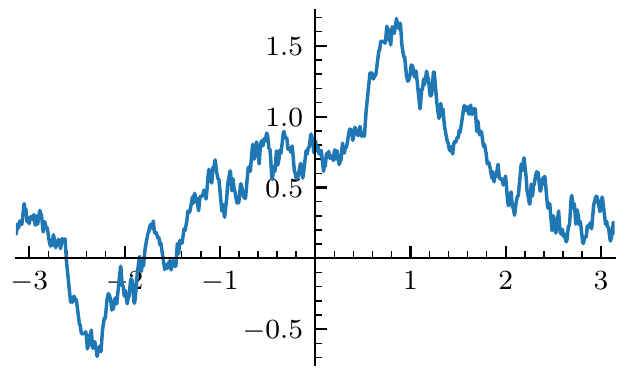}
    FPUT
  \end{minipage}
  \hfil
  \caption{Bi- and uni-directional solution profiles at
    $t=400,\!000$.\label{bu3}}
\end{figure}

However, at long rational times, the solution profiles differ dramatically, as illustrated in \fgas{bu4}{bu5} for two representative such times.  The linearized KdV solution has quantized into a piecewise constant profile, whereas the FPUT system and the Boussinesq models retain a common fractal form.  On the other hand, the latter profiles exhibit an observable adherence to the underlying piecewise constant KdV solution, albeit with a superimposed fractal modulation.  As before, as one increases the number of masses, the relative amplitudes of the fractal parts remain similar, but the magnitude of the frequencies represented in the fractal oscillations increases. 
Recall that the initial data is the discrete Fourier representation \eq{DFTstep} of the step function.  Interestingly, if we use the continuous version \eq{Fstep} instead, which only differs in its higher frequency modes, the graphs do not appreciably change, and so are not displayed.  The only noticeable difference is that, in the latter situation, the piecewise constant KdV profile exhibits a more pronounced Gibbs phenomenon at the discontinuities.

\begin{figure}
  \hfil
  \begin{minipage}[t]{1.75in}
    \centering
    \includegraphics[width=\textwidth]{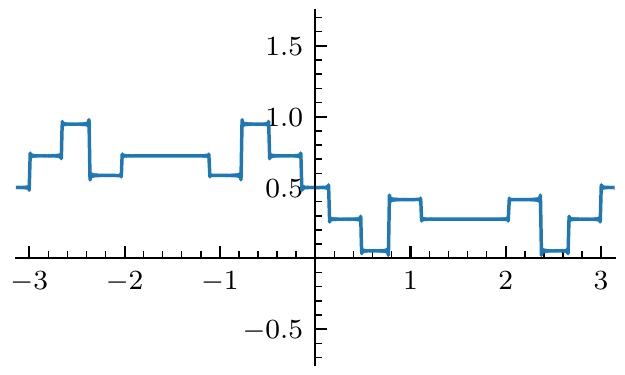}
    \includegraphics[width=\textwidth]{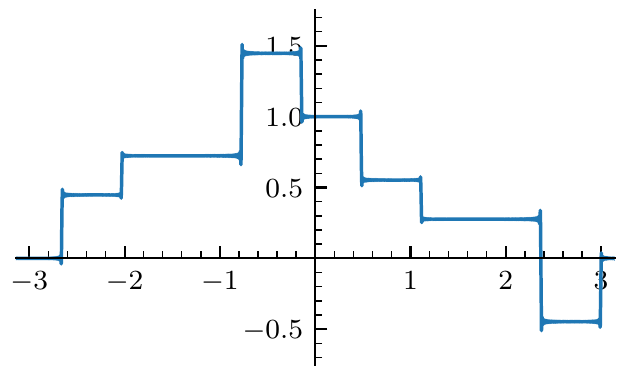}
    KdV
  \end{minipage}
  \hfil
  \begin{minipage}[t]{1.75in}
    \centering
    \includegraphics[width=\textwidth]{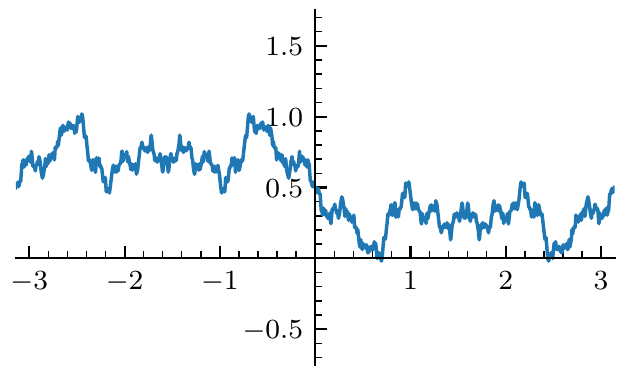}
    \includegraphics[width=\textwidth]{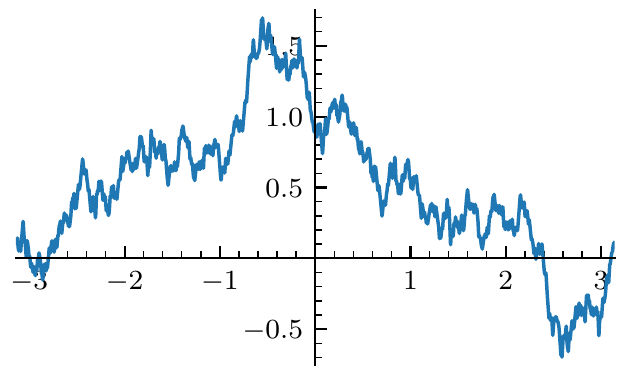}
    Boussinesq
  \end{minipage}
  \hfil
  \begin{minipage}[t]{1.75in}
    \centering
    \includegraphics[width=\textwidth]{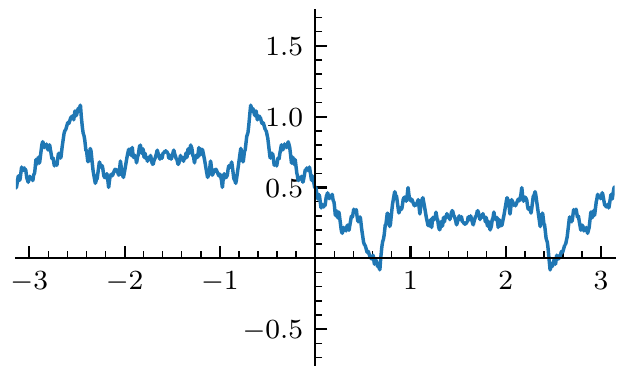}
    \includegraphics[width=\textwidth]{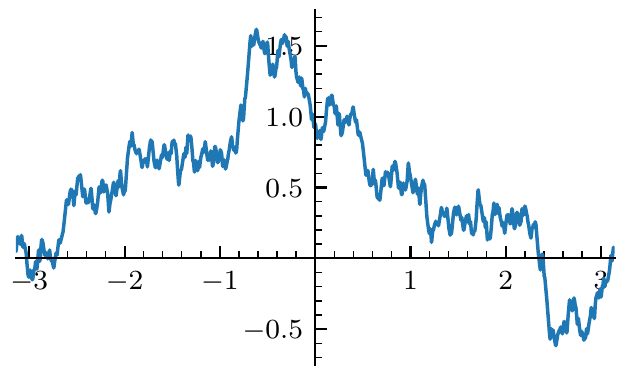}
    FPUT
  \end{minipage}
  \hfil
  \caption{Bi- and uni-directional solution profiles at
    $t=24\pii/(5\:h^2) \approx 400,\!527$.\label{bu4}}
\end{figure}

\begin{figure}
  \hfil
  \begin{minipage}[t]{1.75in}
    \centering
    \includegraphics[width=\textwidth]{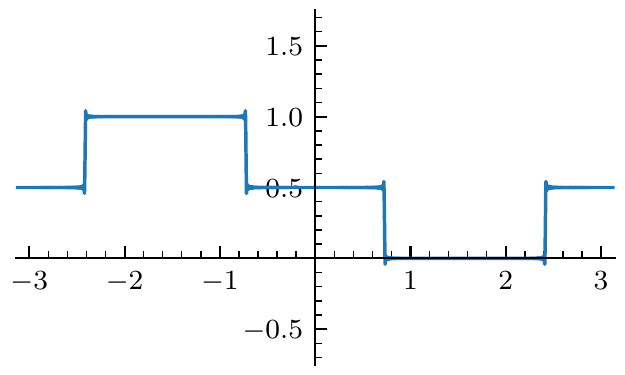}
    \includegraphics[width=\textwidth]{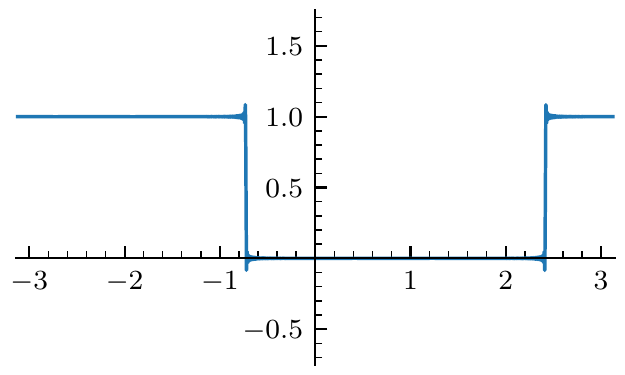}
    KdV
  \end{minipage}
  \hfil
  \begin{minipage}[t]{1.75in}
    \centering
    \includegraphics[width=\textwidth]{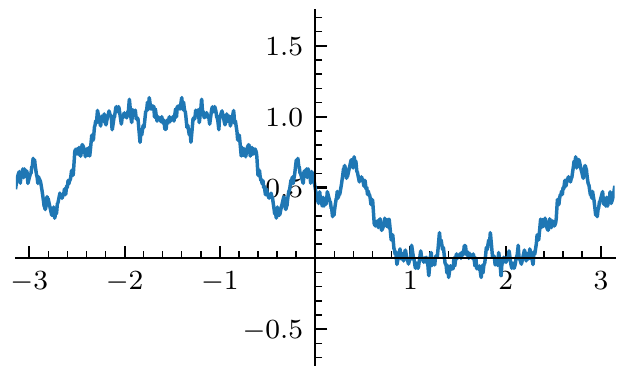}
    \includegraphics[width=\textwidth]{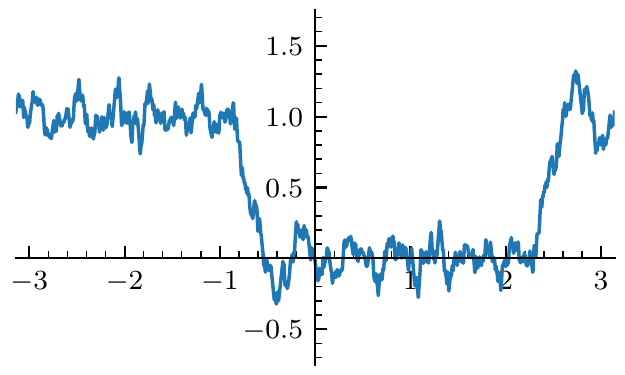}
    Boussinesq
  \end{minipage}
  \hfil
  \begin{minipage}[t]{1.75in}
    \centering
    \includegraphics[width=\textwidth]{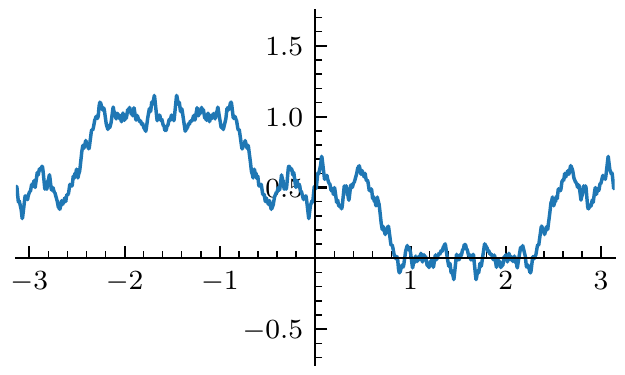}
    \includegraphics[width=\textwidth]{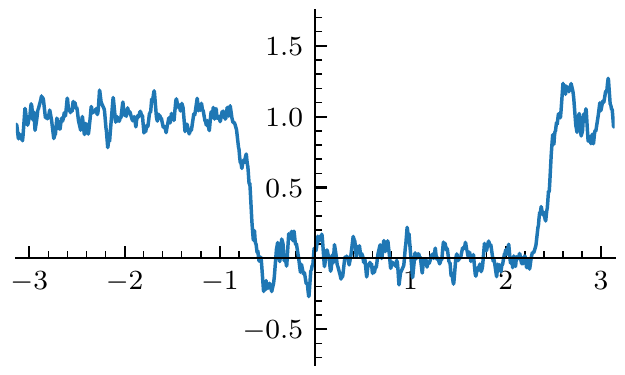}
    FPUT
  \end{minipage}
  \hfil
  \caption{Bi- and uni-directional solution profiles at
    $t=24\pii/h^2$.\label{bu5}}
\end{figure}

\begin{figure}
  \hfil
  \begin{minipage}[t]{1.75in}
    FPUT
    
    \includegraphics[width=\textwidth]{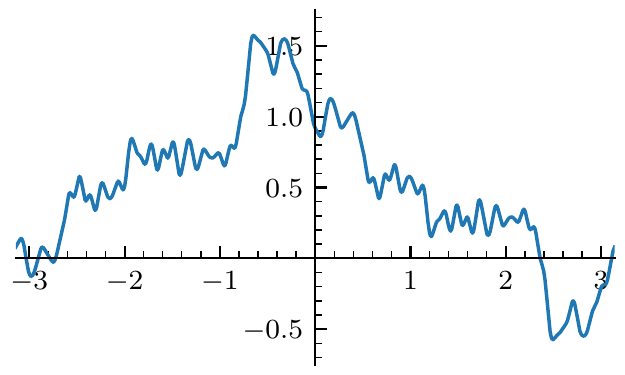}

    \bigskip KdV
    
    \includegraphics[width=\textwidth]{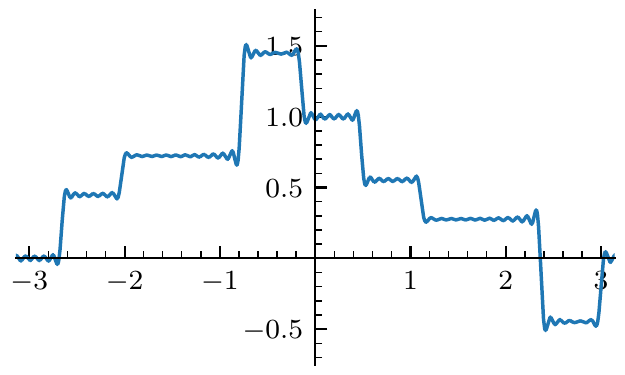}

    \centering $ 1/8 $ of modes
  \end{minipage}
  \hfil
  \begin{minipage}[t]{1.75in}
    \vphantom{FPUT}
    
    \includegraphics[width=\textwidth]{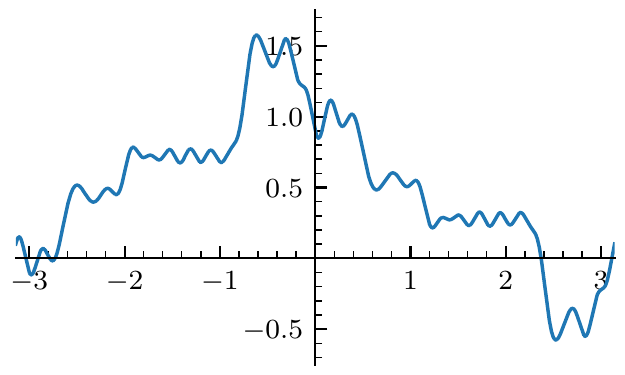}

    \bigskip \vphantom{KdV}
    
    \includegraphics[width=\textwidth]{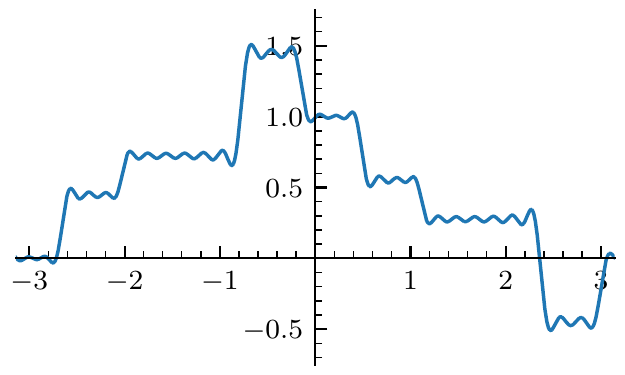}

    \centering $ 1/16 $ of modes
  \end{minipage}
  \hfil
  \begin{minipage}[t]{1.75in}
    \vphantom{FPUT}
    
    \includegraphics[width=\textwidth]{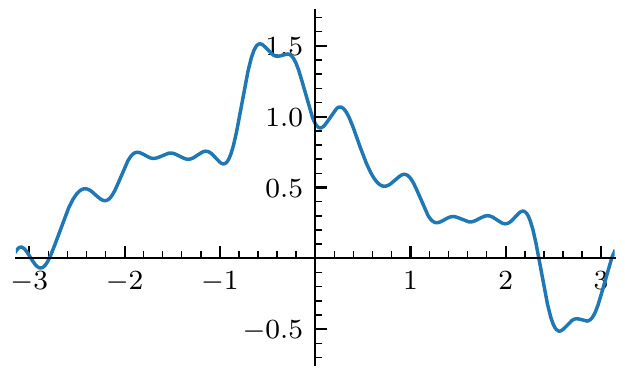}

    \bigskip \vphantom{KdV}
    
    \includegraphics[width=\textwidth]{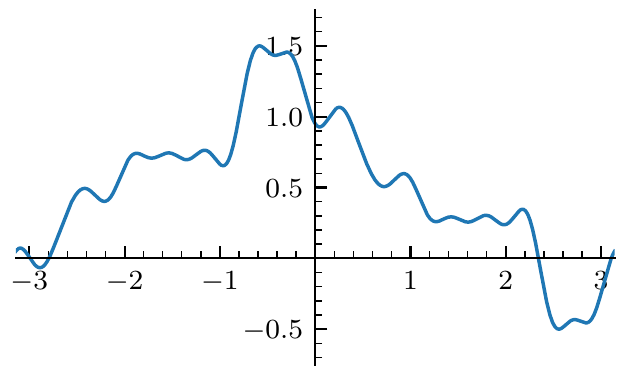}

    \centering $ 1/32 $ of modes
  \end{minipage}
  \hfil
  \caption{Truncated unidirectional solution profiles at $t=24\pii/(5\:h^2) \approx 400,\!527$.\label{tu3}}
\end{figure}

Now, one might argue that the differences between the quantized KdV profiles and the fractal FPUT ones is due to discrepancies in their dispersion relations at the high frequency modes. So let us try eliminating the higher frequency terms by truncating the Fourier sum in order to bring the solutions closer in spirit.  It is surprising that one must eliminate a large majority of the high frequency modes before their respective truncated solution profiles begin to align at the rational quantized times; on the other hand, at the irrational fractalized times they are quite similar no matter how one truncates. 

\begin{figure}
  \hfil
  \begin{minipage}[t]{1.75in}
    FPUT
    
    \includegraphics[width=\textwidth]{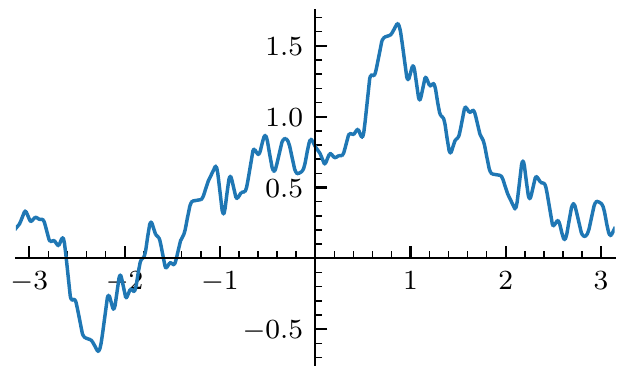}

    \bigskip KdV
    
    \includegraphics[width=\textwidth]{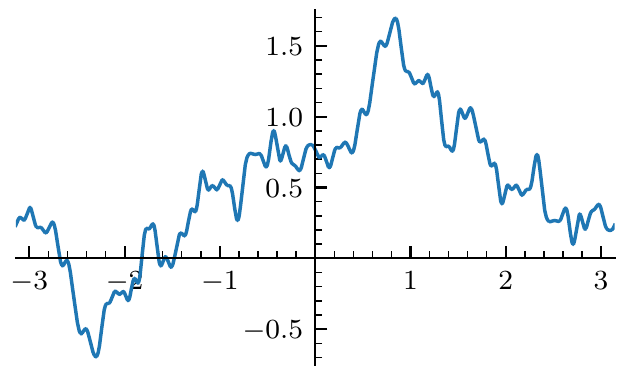}

    \centering $ 1/8 $ of modes
  \end{minipage}
  \hfil
  \begin{minipage}[t]{1.75in}
    \vphantom{FPUT}
    
    \includegraphics[width=\textwidth]{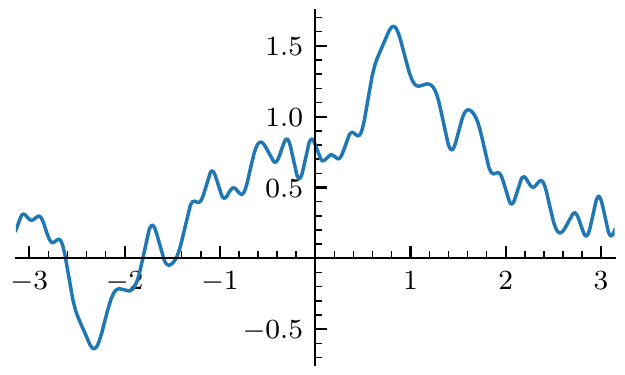}

    \bigskip \vphantom{KdV}
    
    \includegraphics[width=\textwidth]{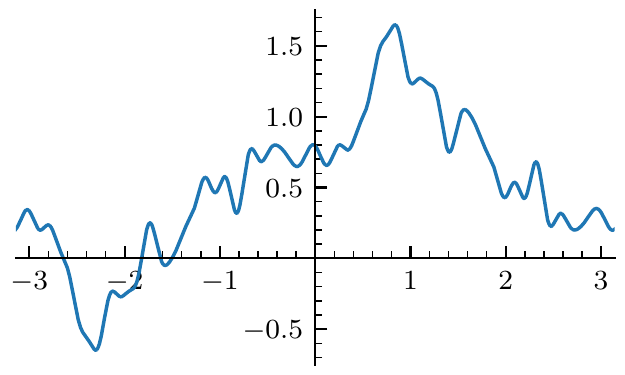}

    \centering $ 1/16 $ of modes
  \end{minipage}
  \hfil
  \begin{minipage}[t]{1.75in}
    \vphantom{FPUT}
    
    \includegraphics[width=\textwidth]{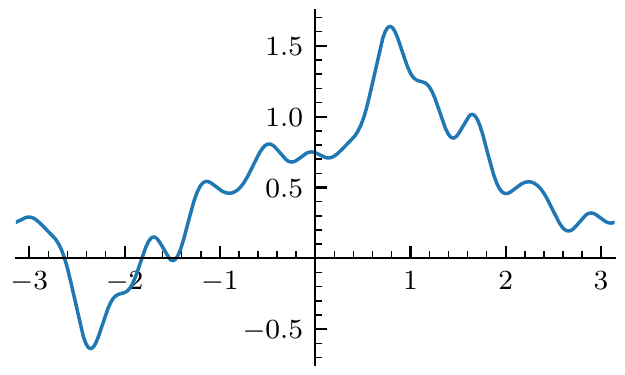}

    \bigskip \vphantom{KdV}
    
    \includegraphics[width=\textwidth]{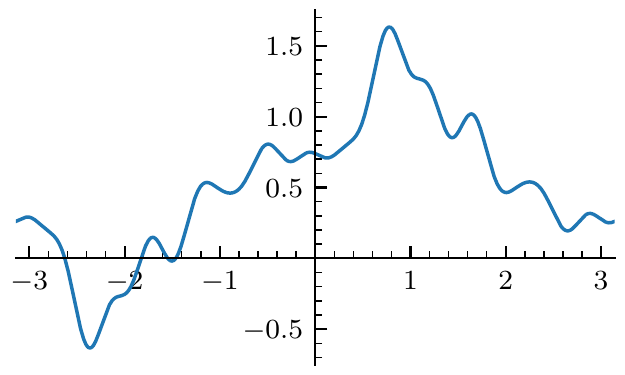}

    \centering $ 1/32 $ of modes
  \end{minipage}
  \hfil
  \caption{Truncated unidirectional solution profiles at
    $t=400,\!000$.\label{tu4}}
\end{figure}

Keeping in mind that we are working with $m=512$ total modes, the first plots in \fg{tu3} are at the same quantized time illustrated in \fg{bu4}, and show the results of summing the first $1/8$, $1/16$, and $ 1/32 $ of the terms in the discrete Fourier summation (i.e., $64$, $32$, and $16$ terms).  The top row shows the resulting truncated profiles for the unidirectional FPUT solution \eq{FPUssR}, while the bottom row shows the corresponding truncated KdV profile \eq{stepsolR}.
Somewhat surprisingly, even retaining $1/8$ of the modes leads to significant differences; these differences persist (albeit more subtly) at the $1/16$ scale, and only at the very coarse $1/32$ scale do they look very close. On the other hand, in \fg{tu4}, which illustrates the same results at the irrational time that were shown in \fg{bu4}, all three pairs of truncated profiles exhibit very similar features, while, as expected, the overall local fractal nature of the profile is curtailed as the number of terms decreases.

Of course, the FPUT mass-spring chain is not a continuum, and so the values of the trigonometric solution \eq{FPUss} --- or its unidirectional counterpart \eq{FPUssR} --- only have physical meaning at the mass nodes.  For the above cases of $\M = 1024$ masses, the differences are imperceptible.  To better illustrate, in \fg{bfk} we plot solution profiles for $\M = 64$ masses, at selected times, comparing the discrete mass displacement profiles, the corresponding continuum FPUT bidirectional solution \eq{FPUss}, and the continuum bidirectional solution \eq{stepsol} with KdV dispersion \eq{lkdvhd}.  Observe that the effects are quite similar to what was presented above, but less pronounced owing to the relatively small number of masses.

Finally, let us investigate whether the Talbot revival phenomenon mentioned in the introduction appears in the FPUT system.  Somewhat surprisingly, given the noticeable effects of dispersive quantization at long rational times, numerical experiments have failed to reveal any observable trace of revival.

\begin{figure}
  \hfil
  \begin{minipage}[t]{1.75in}
    $ t = 1 $
    
    \includegraphics[width=\textwidth]{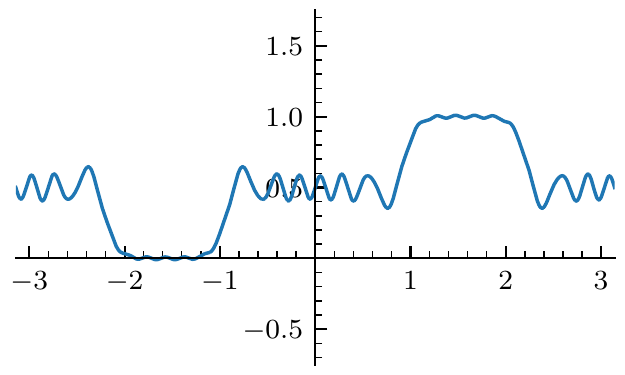}

    \bigskip $ t = 7500 $
    
    \includegraphics[width=\textwidth]{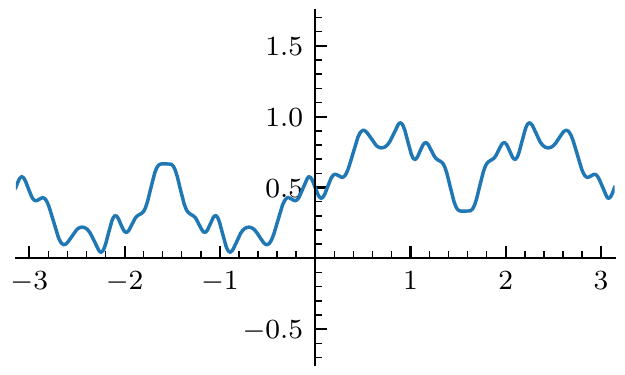}

    \bigskip $t = 24\:\pi/h^2 \approx 7822$
    
    \includegraphics[width=\textwidth]{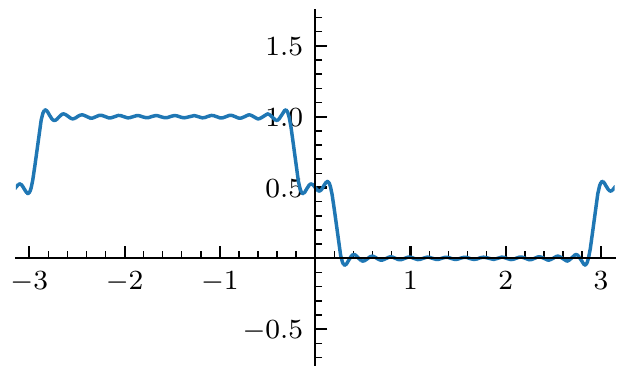}
    
    \centering KdV
  \end{minipage}
  \hfil
  \begin{minipage}[t]{1.75in}
    \vphantom{$ t = 1 $}
    
    \includegraphics[width=\textwidth]{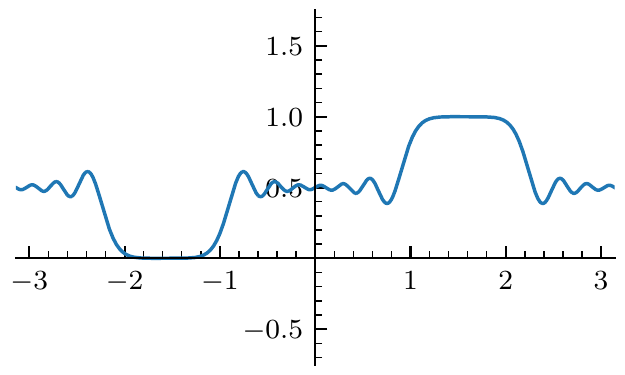}

    \bigskip \vphantom{$ t = 7500 $}

    \includegraphics[width=\textwidth]{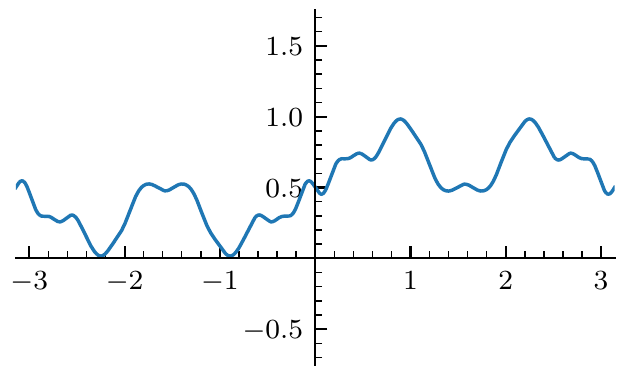}

    \bigskip \vphantom{$t = 24\:\pi/h^2 \approx 7822$}

    \includegraphics[width=\textwidth]{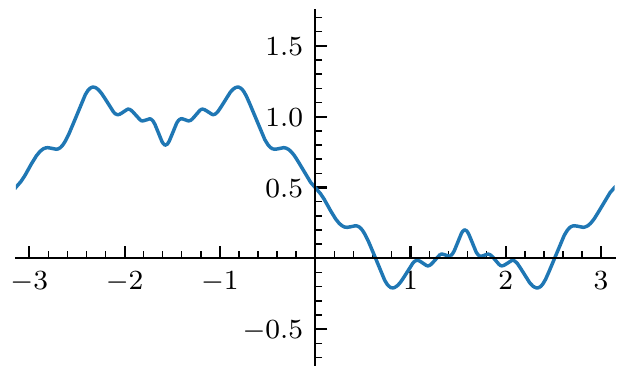}

    \centering
    Continuum FPUT
  \end{minipage}
  \hfil
  \begin{minipage}[t]{1.75in}
    \vphantom{$ t = 1 $}

    \includegraphics[width=\textwidth]{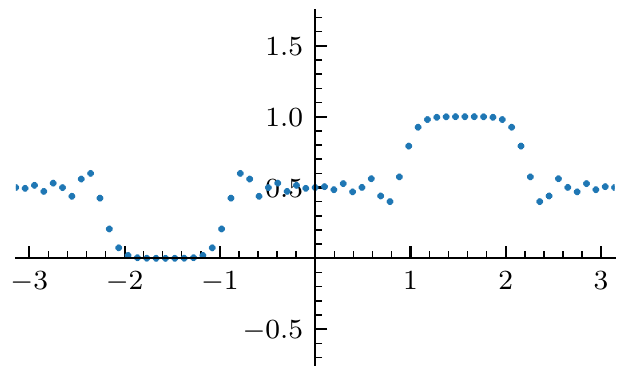}

    \bigskip \vphantom{$ t = 7500 $}

    \includegraphics[width=\textwidth]{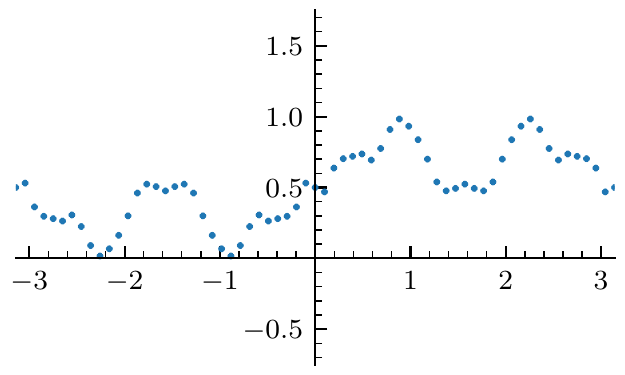}

    \bigskip \vphantom{$t = 24\:\pi/h^2 \approx 7822$}

    \includegraphics[width=\textwidth]{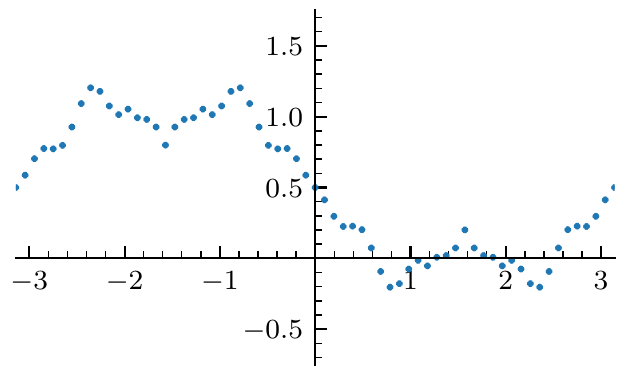}

    \centering
    Discrete FPUT
  \end{minipage}
  \hfil
  \caption{Bidirectional solution profiles for the discrete and continuum FPUT system and the KdV model with $m=32$.\label{bfk}}
\end{figure}

To model the delta function initial displacement, we displace the center mass by a unit\fnote{Since we are dealing with a linear system, the magnitude of the displacement of the single mass does not, modulo rescaling, affect the response.}. (This is equivalent to equipartitioning the initial energy into all the Fourier modes.)  
\fg{rfk} plots the resulting solutions, in the case of  $\M = 64$ masses, whose initial data is the (Fourier series for) the delta function, at the indicated long rational times.  These, as always, are obtained by explicitly summing over the first $\m = 32$ modes.  Keep in mind that the Fourier series of the delta function and resulting fundamental solution to the continuum model is highly oscillatory, and only converges weakly to the distributional revival profile, consisting of a finite linear combination of delta functions, at rational times.  The first column plots the solutions to the bidirectional KdV model; the discrete oscillatory peaks indicate the appearance of a revival.  The second column plots the corresponding FPUT solution; here, there is no appreciable sign of concentration of the solution profiles and hence no apparent revival.  Similar behaviors have been observed at other (long) times, with differing number of masses.  The KdV profiles are fractal at irrational times and concentrated in accordance with a revival at rational times, whereas the FPUT profiles are more or less uniformly oscillatory at all times.  Thus, the Korteweg--deVries equation appears to do a poor job modeling the behavior of the FPUT system at these particular times.

\begin{figure}
  \hfil
  \begin{minipage}[t]{1.0in}
    \vspace{-38pt}
    $t = 24\:\pi/(5\:h^2)$

    \vspace{64pt}
    $t = 24\:\pi/h^2$
  \end{minipage}
  \hfil
  \begin{minipage}[t]{1.75in}
    \centering
    \includegraphics[width=\textwidth]{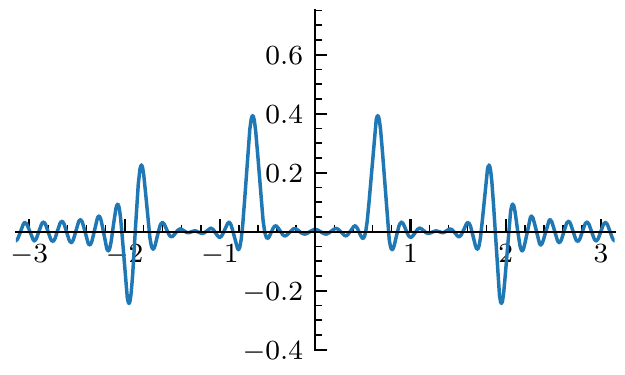}
    \includegraphics[width=\textwidth]{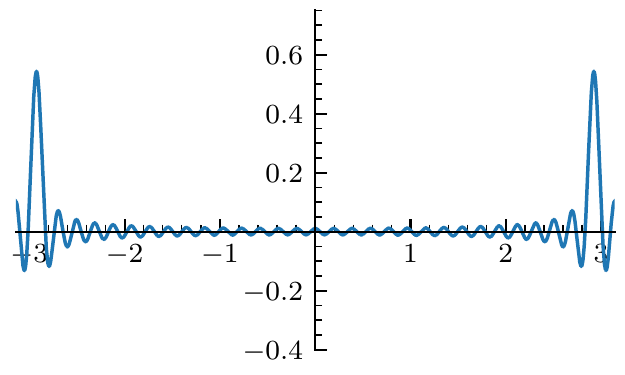}
    KdV
  \end{minipage}
  \hfil
  \begin{minipage}[t]{1.75in}
    \centering
    \includegraphics[width=\textwidth]{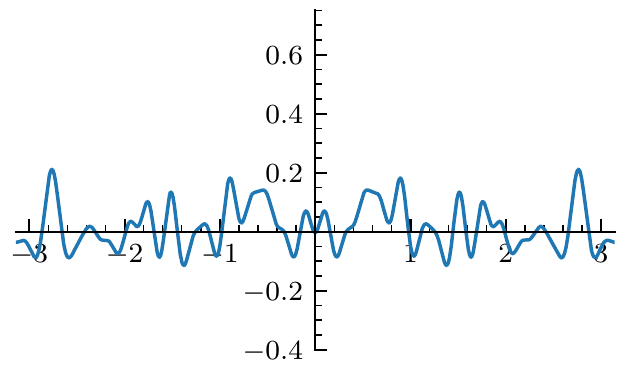}
    \includegraphics[width=\textwidth]{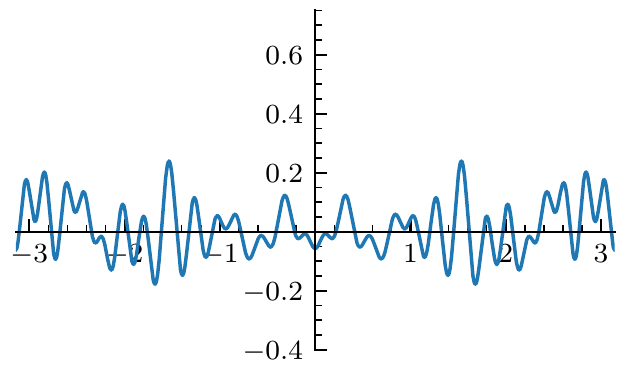}
    FPUT
  \end{minipage}
  \hfil
  \begin{minipage}[t]{0.5in}
    \hspace{0pt}
  \end{minipage}
  \hfil
  \caption{Revival and lack thereof.\label{rfk}}
\end{figure}

\Section n Numerical Investigation of Nonlinear FPUT Chains.

For the linear problems featured in the previous sections, Fourier
series techniques make it possible to compute exact solutions (up to
floating point error) for extremely long times at rather low
computational expense---essentially the cost of a Fast Fourier
Transform (FFT). However, once nonlinearity is introduced, the
resulting systems of differential equations can usually only be solved
approximately, by using a numerical integrator with sufficiently small
time step size. Since the FPUT system is Hamiltonian, it is natural to
consider the class of symplectic integrators, which have excellent
long-time numerical properties for Hamiltonian systems, including
near-conservation of energy and slow growth of global error, \rf{HLW}.

There has been substantial investigation of the application of
symplectic integrators (and other geometric numerical integrators) to
the variant of the FPUT problem appearing in \rf{GGMV}, with
alternating stiff linear and soft nonlinear springs. Much of this work
has focused on methods, such as trigonometric and modified
trigonometric integrators, which can take large time steps in order to
simulate the slow-scale nonlinear dynamics without needing to resolve
the fast-scale linear oscillations. See \rf{HLW; Chapter XIII} for a
survey and \rf{StGr,McSt} for more recent work on modified
trigonometric integrators, including the IMEX method.

However, the phenomenon of dispersive quantization is quite delicate
and requires the accurate resolution of high wave number oscillations,
\rf{COdisp}. To illustrate the challenge this poses, we begin by
considering the application of two widely-used symplectic integrators,
the explicit St\"ormer/Verlet method and implicit midpoint method, to
the harmonic oscillator $ \ddot{y} = - \omega ^2 y $,
$ \omega \geq 0 $. Let $ y ^j \approx y ( j \Delta t ) $ denote the
approximation produced by the method at the $j$th time step, where
$ \Delta t $ is the time step size. The St\"ormer/Verlet method gives
\begin{equation}
  y ^{ j + 1 } - 2 y ^j + y ^{ j -1 } = - (\omega \:\Delta t) ^2 y ^j ,
\end{equation}
and substituting the exponential ansatz $ y ^j = e ^{\i \widetilde{ \omega } j \Delta t } $ yields, by a similar calculation to \eqr{dFPU}{FPUd},
\begin{equation}
  \sin ^2 \f2 \widetilde{ \omega } \Delta t = \bpa{\f2 \omega \Delta t} ^2 \ \Longrightarrow \ \widetilde{ \omega } = \frac2{\Delta t} \arcsin \f2 \:\omega \:\Delta t = \omega \Bigl( 1 + \f{24} ( \omega \:\Delta t ) ^2 + \ro O \bigl( (\omega \:\Delta t ) ^4 \bigr) \Bigr) .
\end{equation}
Hence, the St\"ormer/Verlet method produces harmonic oscillations with
modified frequency $ \widetilde{ \omega } $. Note that
$ \lvert \f2 \omega \Delta t \rvert \leq 1 $ is necessary for
$ \widetilde{ \omega } $ to be defined, and this is precisely the
linear stability condition for St\"ormer/Verlet. Similarly, the
midpoint method gives,
\begin{equation}
  y ^{ j + 1 } - 2 y ^j + y ^{ j -1 } = - \,\bpa{ \f2\: \omega\: \Delta t }^2 ( y ^{ j + 1 } + 2\: y ^j + y ^{ j -1 } ),
\end{equation}
and substituting the exponential ansatz yields
\begin{equation}
  \tan ^2 \f2 \widetilde{ \omega } \Delta t = (\f2 \: \omega\: \Delta t) ^2 \ \Longrightarrow \ \widetilde{ \omega } = \frac2{\Delta t} \arctan \f2 \:\omega\: \Delta t = \omega \Bigl( 1 - \f{12} ( \omega \:\Delta t ) ^2 + \ro O \bigl( (\omega \:\Delta t ) ^4 \bigr) \Bigr) .
\end{equation}
In contrast to St\"ormer/Verlet, this modified frequency is defined
without restrictions on $ \omega \Delta t $, which reflects the
unconditional linear stability of the midpoint method.

\begin{figure}
  \hfil
  \begin{minipage}[t]{1.75in}
    \centering
    \includegraphics[width=\textwidth]{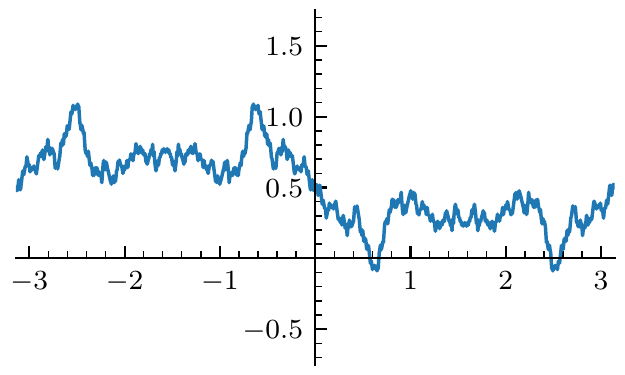}
    \includegraphics[width=\textwidth]{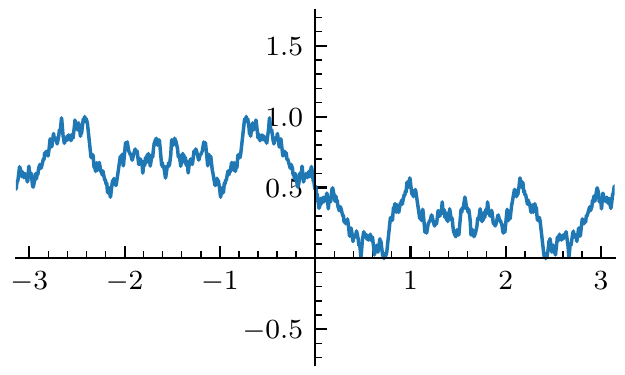}
    $ \Delta t = 10 ^{ - 4 } $
  \end{minipage}
  \hfil
  \begin{minipage}[t]{1.75in}
    \centering
    \includegraphics[width=\textwidth]{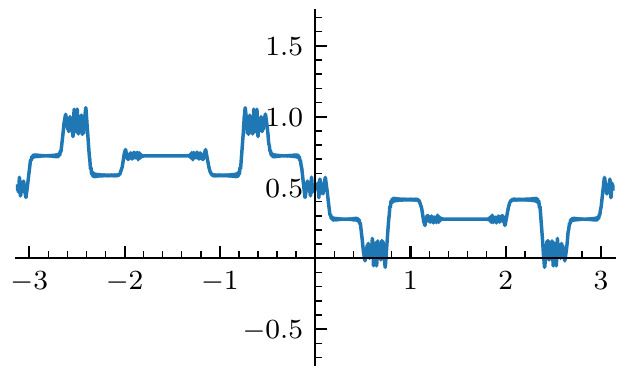}
    \includegraphics[width=\textwidth]{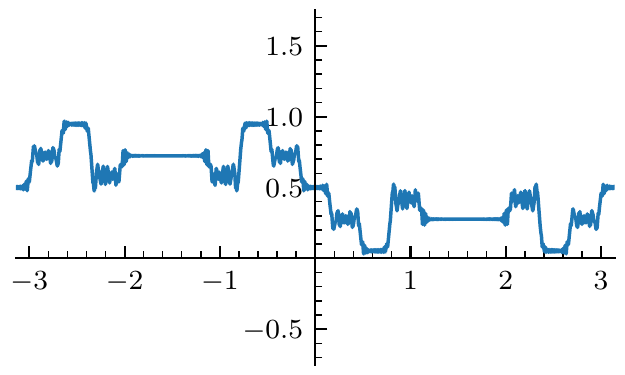}
    $ \Delta t = 10 ^{ - 5 } $
  \end{minipage}
  \hfil
  \begin{minipage}[t]{1.75in}
    \centering
    \includegraphics[width=\textwidth]{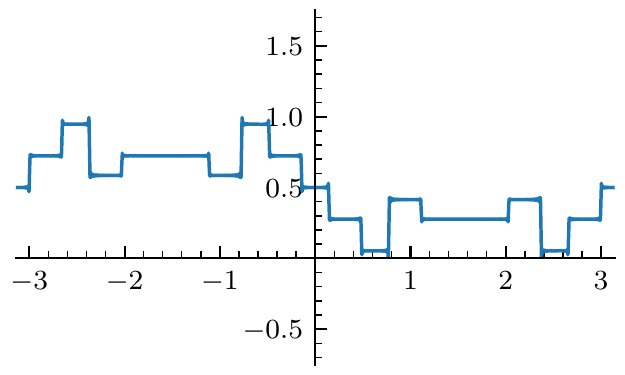}
    \includegraphics[width=\textwidth]{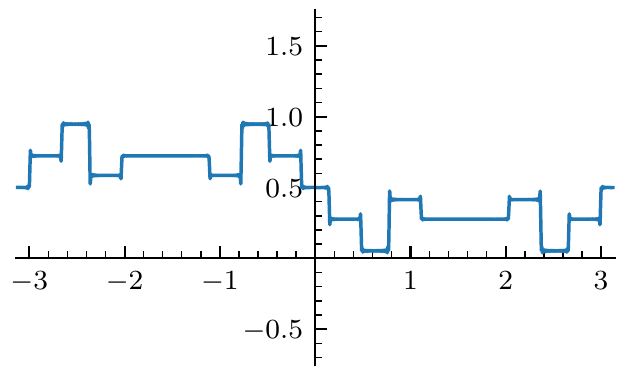}
    $ \Delta t = 10 ^{ - 6 } $
  \end{minipage}
  \hfil
  \caption{Numerical approximation of the bidirectional KdV solution
    profile with $ m = 512 $ at $t=24\pii/(5\:h^2)$, showing the
    effect of time step size $ \Delta t $ for the St\"ormer/Verlet
    method (top) and midpoint method (bottom).\label{kdv_integrators}
  }
\end{figure}

\begin{figure}
  \hfil
  \begin{minipage}[t]{1.75in}
    \centering
    \includegraphics[width=\textwidth]{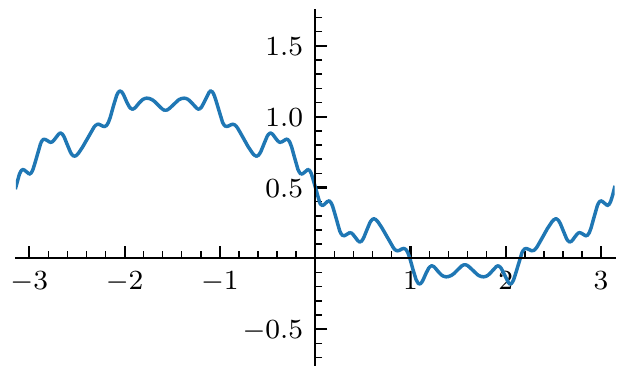}
    \includegraphics[width=\textwidth]{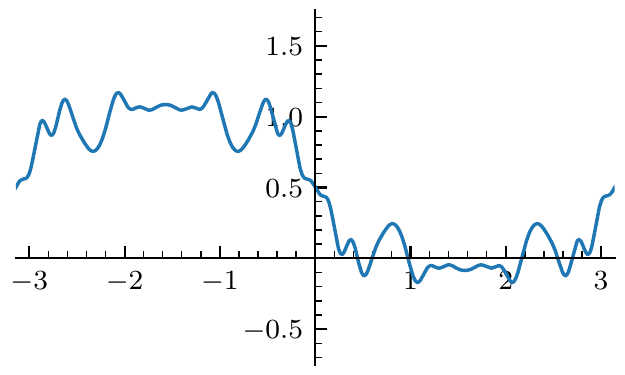}
    $ \Delta t = 10 ^{ - 2 } $
  \end{minipage}
  \hfil
  \begin{minipage}[t]{1.75in}
    \centering
    \includegraphics[width=\textwidth]{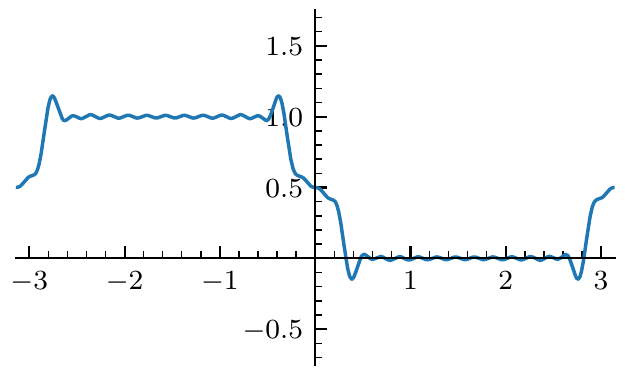}
    \includegraphics[width=\textwidth]{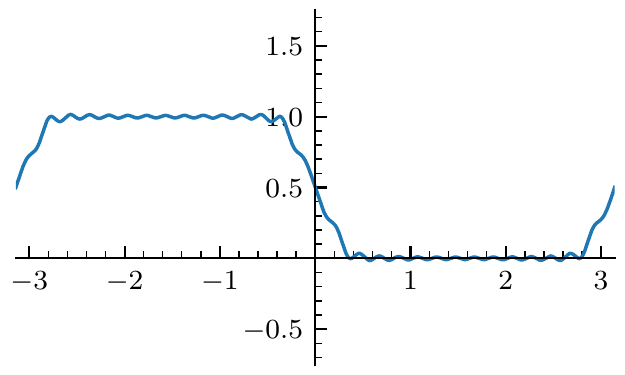}
    $ \Delta t = 10 ^{ - 3 } $
  \end{minipage}
  \hfil
  \begin{minipage}[t]{1.75in}
    \centering
    \includegraphics[width=\textwidth]{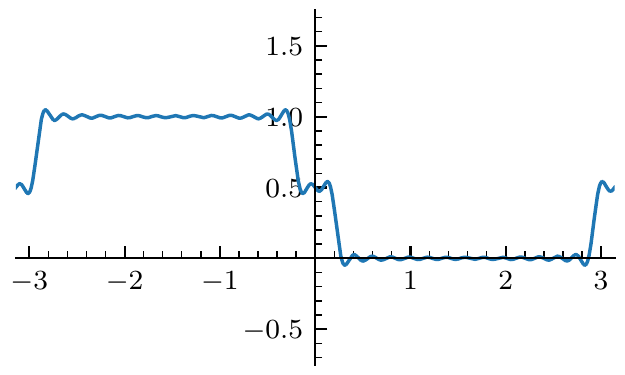}
    \includegraphics[width=\textwidth]{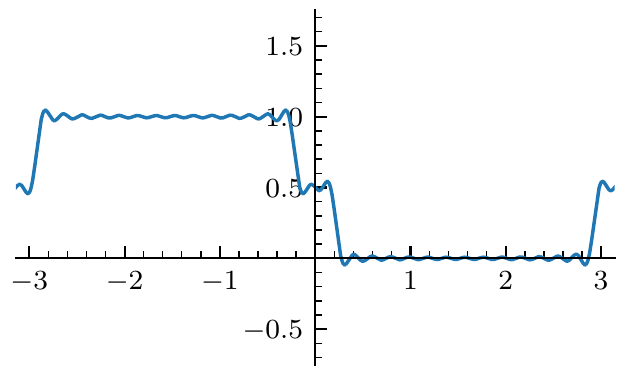}
    $ \Delta t = 10 ^{ - 4 } $
  \end{minipage}
  \hfil
  \caption{Numerical approximation of the bidirectional KdV solution
    profile with $ m = 32 $ at $t=24\pii/h^2$, showing the effect of
    time step size $ \Delta t $ for the St\"ormer/Verlet method (top)
    and midpoint method (bottom).\label{kdv_integrators_32} }
\end{figure}

\fg{kdv_integrators} illustrates the effect of replacing $\omega$ by
the modified frequencies $ \widetilde{ \omega } $ in the dispersion
relation for the bidirectional KdV model with $ m = 512 $ and
$ t = 24\pii/(5\:h^2) $, showing that quantization does not become
visible unless $ \Delta t $ is very small, much smaller than needed
for numerical stability. At $ \Delta t = 10 ^{ - 4 } $, the solution
profile is qualitatively indistinguishable from the fractal profiles
of the FPUT and Boussinesq models in \fg{bu4}. The first hints of
quantization are visible at $ \Delta t = 10 ^{ - 5 } $, and only by
$ \Delta t = 10 ^{ - 6 } $ does the solution appear to have converged
sufficiently to the true, quantized KdV profile. Since
$ t \approx 4 \cdot 10 ^{ 5 } $ such a simulation would require on the
order of $ 10 ^{ 11 } $ time steps to observe quantization, even
before the effects of nonlinearity are taken into
account. \fg{kdv_integrators_32} repeats this experiment for a shorter
chain with $ m = 32 $ and $ t = 24\pii/h ^2 \approx 8 \cdot 10 ^3 $
(compare \fg{bfk}), where quantization occurs earlier and the time
step size restriction is less severe, requiring on the order of
$ 10 ^7 $ steps.

To overcome the computational obstacle of small step size, we turn to
higher-order Hamiltonian splitting methods, \crf{HLW,McQu}, which
converge more quickly as $ \Delta t \rightarrow 0 $ while still
preserving symplectic structure. Write the FPUT system \eq{rFPU} in
the first-order form
\begin{equation}
  \begin{aligned}
    \dot{ u } _n &= v _n ,\\
    \dot{ v } _n &= \frac{ c ^2 }{ h ^2 } \bigl[ F ( u _{ n + 1 } - u _n ) - F ( u _n - u _{ n -1 } ) \bigr] ,
  \end{aligned} 
\end{equation}
where $ F (y) = y + N (y) $. There are two natural ways to split this
into two Hamiltonian systems, each of which can be integrated
exactly. The first is
\begin{equation}
    \label{drift_force_splittingeq*}
  \begin{alignedat}{2}
    \dot{ u } _n &= v _n , \qquad \qquad  &   \dot{ u } _n &= 0 ,\\
    \dot{ v } _n &= 0, \qquad \qquad & \dot{ v } _n &= \frac{ c ^2 }{ h ^2 } \bigl[
    F ( u _{ n + 1 } - u _n ) - F ( u _n - u _{ n -1 } ) \bigr],
  \end{alignedat}
\end{equation}
where each of these can be integrated exactly since $v$ is constant in
the first system and $u$ is constant in the second. The second
splitting is
\begin{equation}
    \label{linear_nonlinear_splittingeq*}
  \begin{alignedat}{2}
    \dot{ u } _n &= v _n , \qquad \qquad  &   \dot{ u } _n &= 0 ,\\
    \dot{ v } _n &= \frac{ c ^2 }{ h ^2 } ( u _{ n + 1 } - 2 u _n + u
    _{ n -1 } ), \qquad \qquad & \dot{ v } _n &= \frac{ c ^2 }{ h ^2 }
    \bigl[ N ( u _{ n + 1 } - u _n ) - N ( u _n - u _{ n -1 } ) \bigr],
  \end{alignedat}
\end{equation}
where the first system is simply linear FPUT, which can be integrated
exactly using the Fourier series techniques applied
previously. Splitting methods approximate the time-$ \Delta t $ flow
of the full system by alternating between the flows of the two
subsystems, which we denote by $ \varphi ^A _{ a _i \Delta t } $ and
$ \varphi ^B _{ b _i \Delta t } $, where
$ \sum _i a _i = \sum _i b _i = 1 $.

One of the simplest splitting methods is
$ \varphi ^B _{ \Delta t/2 } \circ \varphi ^A _{ \Delta t } \circ
\varphi ^B _{ \Delta t/2 } $, called \emph{Strang
  splitting},~\rf{Strang}. For the splitting
\eq{drift_force_splitting}, this results in a first-order formulation
of the St\"ormer/Verlet method, sometimes called \emph{velocity
  Verlet}. Alternatively, for the splitting
\eq{linear_nonlinear_splitting}, where the linear flow
$ \varphi _{ \Delta t } ^A $ is computed using Fourier series
techniques, the Strang splitting gives the so-called \emph{split-step
  Fourier method}. Whenever the composition of flows is symmetric, the
resulting method has even order, and in particular, the Strang
splitting is order-$2$.

\begin{table}
  \centering
  \begin{tabular}{|lrccc|}
    \hline
    && $ \Delta t = 10 ^{ - 2 } $ & $ \Delta t = 10 ^{ - 3 } $ & $ \Delta t = 10 ^{ - 4 } $ \\
    \hline
    \multirow{2}{*}{$ \alpha = 0.005 $} &
    St\"ormer/Verlet & 4.347e-03 & 4.332e-05 & 4.331e-07 \\
    & Split-Step Fourier & 1.423e-06 & 1.421e-08 & 1.474e-10 \\
    \hline 
    \multirow{2}{*}{$ \alpha = 0.05 $} &   
    St\"ormer/Verlet & 4.348e-03 & 4.333e-05 & 4.332e-07 \\
    & Split-Step Fourier & 1.429e-05 & 1.427e-07 & 1.427e-09 \\
    \hline
    \multirow{2}{*}{$ \alpha = 0.5 $} &   
    St\"ormer/Verlet & 4.544e-03 & 4.529e-05 & 4.529e-07 \\
    & Split-Step Fourier & 1.895e-04 & 1.891e-06 & 1.891e-08\\
    \hline
  \end{tabular}
  \bigskip
  \caption{Numerical error of Strang splitting methods at $ t = 1 $
    for the FPUT model with $ m = 32 $ and quadratic nonlinearity
    $ N (y) = \alpha y ^2 $.\label{table:splitting_error}}
\end{table}

For the linear FPUT system, the split-step Fourier method gives the
exact solution in a single step. However, once nonlinearity is
introduced, we find that the split-step Fourier method has only modest
benefits over St\"ormer/Verlet, which are outweighed in practice by
the additional cost of performing an FFT and inverse FFT at every
step. Table~\ref{table:splitting_error} illustrates this for the
quadratic nonlinearity $ N = \alpha y ^2 $, showing that the error in
St\"ormer/Verlet is $ \ro O \bigl( ( \Delta t ) ^2 \bigr) $ uniformly
as $\alpha \rightarrow 0 $, while the split-step Fourier method is
$ \ro O \bigl( \alpha ( \Delta t ) ^2 \bigr) $. Therefore, we focus
our attention on methods based on the splitting
\eq{drift_force_splitting}, whose steps are less expensive to compute
since they do not require transforming to Fourier space and back.

\emph{Runge--Kutta--Nystr\"om} (RKN) methods are designed specifically
for splittings of the form \eq{drift_force_splitting}, i.e., for
second-order Newtonian systems written in first-order form using a
velocity variable. Of these, we chose the optimal 14-stage order-6 RKN
method of Blanes and Moan, \rf{BlMo}, which has the symmetric form
\begin{equation}
  \varphi _{ a _1 \Delta t } ^A \circ \varphi _{ b _1 \Delta t } ^B \circ \cdots \circ \varphi _{ a _7 \Delta t } ^A \circ \varphi _{ b _7 \Delta t } ^B \circ \varphi _{ a _8 \Delta t } ^A \circ \varphi _{ b _7 \Delta t } ^B \circ \varphi _{ a _7 \Delta t } ^A \circ \cdots \circ  \varphi _{ b _1 \Delta t } ^B \circ \varphi _{ a _1 \Delta t } ^A ,
\end{equation}
where the coefficients $ a _i $, $ b _i $ are
\begin{equation}
  \begin{alignedat}{2}
    a _1 &= 0.0378593198406116,\qquad  & b _1 &= 0.09171915262446165, \\
    a _2 &= 0.102635633102435,\qquad  & b _2 &= 0.183983170005006, \\
    a _3 &= -0.0258678882665587,\qquad  & b _3 &= -0.05653436583288827, \\
    a _4 &= 0.314241403071447,\qquad  & b _4 &= 0.004914688774712854, \\
    a _5 &= -0.130144459517415,\qquad  & b _5 &= 0.143761127168358, \\
    a _6 &= 0.106417700369543,\qquad  & b _6 &= 0.328567693746804, \\
    a _7 &= -0.00879424312851058,\qquad  & b _7 &= \f2 - ( b _1 + \cdots + b _6 ), \\
    a _8 &= 1 - 2 ( a _1 + \cdots + a _7 ) .
  \end{alignedat}
\end{equation}
Blanes and Moan calculated these coefficients to minimize the constant
in the order-$6$ error estimate. Although a step of this method is 14
times as expensive as a step of St\"ormer/Verlet, it requires vastly
fewer steps owing to its faster convergence. For the linear FPUT
problem with $ m = 32 $ and $ t = 24\pii/h ^2 $, we observe that the
St\"ormer/Verlet method gives an error on the order of $ 10 ^{ - 3 } $
for $ 10 ^8 $ steps and $ 10 ^{ - 5 } $ for $ 10 ^9 $ steps; by
contrast, the RKN method gives an error on the order of
$ 10 ^{ - 3 } $ for only $ 10 ^5 $ steps and $ 10 ^{ - 9 } $ for
$ 10 ^6 $ steps.

\begin{figure}
  \hfil
  \begin{minipage}[t]{1.75in}
    $ t = 7500 $
    
    \includegraphics[width=\textwidth]{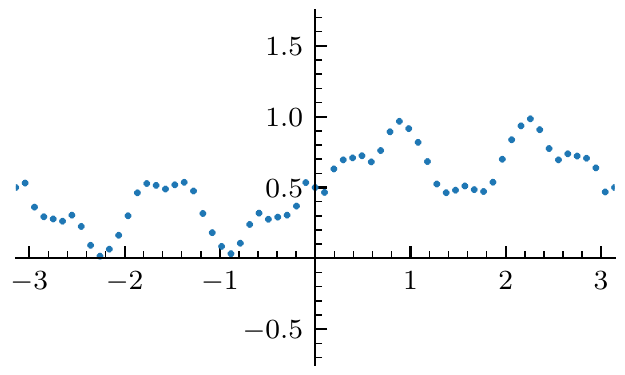}

    \bigskip $t = 24\:\pi/h^2 \approx 7822$
    
    \includegraphics[width=\textwidth]{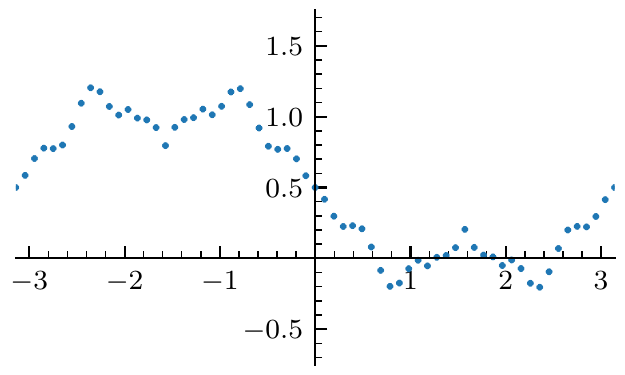}

    \centering $ \alpha = 0.005 $
  \end{minipage}
  \hfil
  \begin{minipage}[t]{1.75in}
    \vphantom{$ t = 7500 $}

    \includegraphics[width=\textwidth]{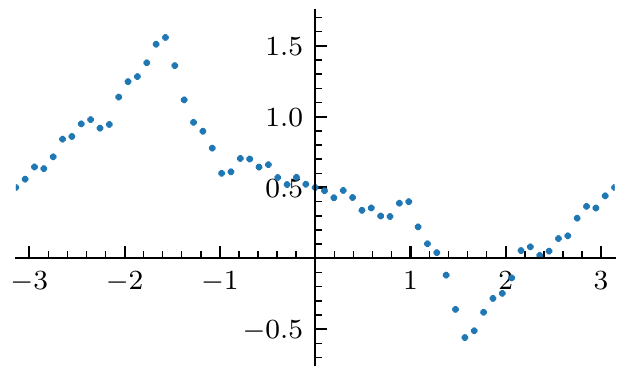}

    \bigskip \vphantom{$t = 24\:\pi/h^2 \approx 7822$}
    
    \includegraphics[width=\textwidth]{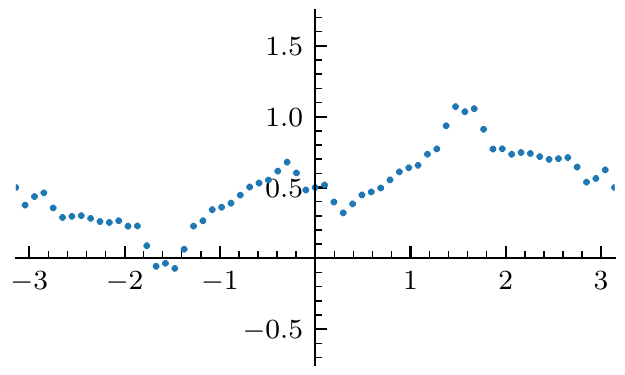}

    \centering $ \alpha = 0.05 $
  \end{minipage}
  \hfil
  \begin{minipage}[t]{1.75in}
    \vphantom{$ t = 7500 $}

    \includegraphics[width=\textwidth]{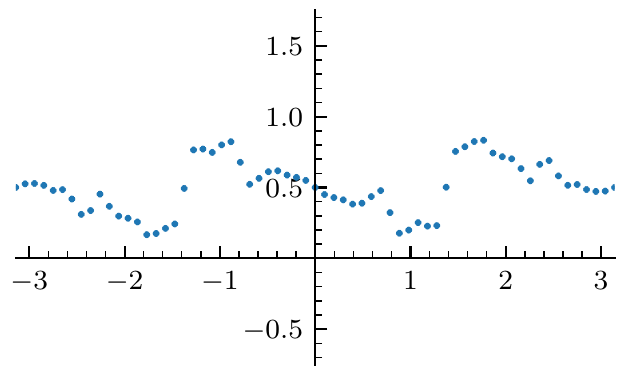}

    \bigskip \vphantom{$t = 24\:\pi/h^2 \approx 7822$}

    \includegraphics[width=\textwidth]{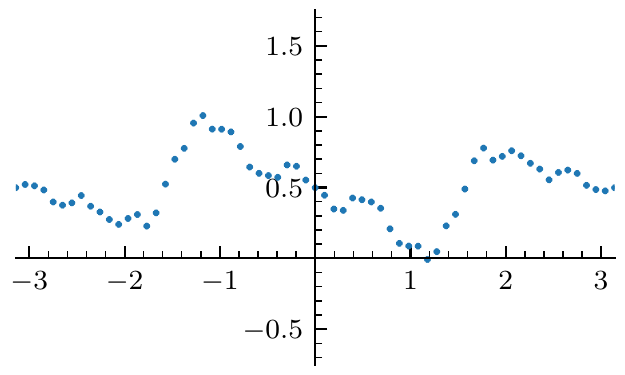}

    \centering $ \alpha = 0.5 $
  \end{minipage}
  \hfil
  \caption{Bidirectional solution profiles for the discrete FPUT
    system with $ m = 32 $ and quadratic nonlinearity
    $ N (y) = \alpha y ^2 $.\label{qfput}}
\end{figure}

\fg{qfput} shows discrete solution profiles at $ t = 7500 $ and
$ t = 24\pii/h ^2 \approx 7822 $ for the FPUT model with $ m = 32 $
and quadratic nonlinearity $ N (y) = \alpha y ^2 $, computed using the
Blanes--Moan RKN method with $ 10 ^6 $ time steps. For
$ \alpha = 0.005 $, the solutions are nearly identical to the linear
discrete FPUT profiles observed in \fg{bfk}. As the strength of the
nonlinearity increases, we observe noticeably different profiles at
$ \alpha = 0.05 $ and $ 0.5 $. However, unlike with the KdV model, we
still do not observe any dispersive quantization, and there does not
appear to be any qualitative difference between the profiles at
$ t = 7500 $ and $ t = 24\pii/h ^2 $, just as with the linear FPUT
model.

 In summary, owing to lack of sufficient computational power to increase the number of masses --- which simultaneously requires extending the time interval of interest --- we are as yet unable to definitely say to what extent dispersive fractalization and quantization appears in nonlinear periodic FPUT chains.  Nevertheless, we feel reasonably confident  in stating that there is a noticeable effect, that is certainly worthy of further investigation. Our claim is bolstered by the appearance of these phenomena in a range of nonlinear model partial differential equations, \rf{COdisp} with rigorous estimates of the fractal dimension of the profiles at irrational times provided in \rf{ChErTz, ErSh}, most relevantly the Korteweg--deVries equation which arises as a continuum model for the quadratic FPUT system and the modified Korteweg-deVries equation, which arises for the cubic version.  

\Section x Discussion.

In conclusion, we have shown that the solution to the periodic linear Fermi--Pasta--Ulam--Tsingou chain, with a step function as initial displacement and zero initial velocity, exhibits a fractal-like solution profile at large times, namely $t = \Oh{-2} = \O{\M^2}$, where $\M$ is the number of masses, and $h$ their spacing around the unit circle.  Of course, being purely discrete, the solution cannot be genuinely fractal, even when extended into a continuous trigonometric interpolating function, because it only involves a sum over a finite number of Fourier modes.  Moreover, it does not become fractal in the final $h \to 0$ limit since the limiting equation is merely the very basic linear second order wave equation \eq{w2}, whose solution is a combination of traveling waves, and hence piecewise constant at all times. Indeed, as $h \to 0$, all of the observed behavior on medium and long time scales moves off to infinity, and the solution converges (weakly) to the corresponding  solution to the simple limiting wave equation, with all times now being classified as ``short''. On the other hand, all of the regularized bidirectional continuum models have genuinely fractal solutions at a dense set of times, which closely follow the FPUT solution at the given resolution.  In contrast, the bi- and uni-directional Korteweg--deVries models mimic the FPUT and Boussinesq solutions at irrational times, but exhibit a very different dispersive quantization profile at rational times. Be that as it may, the latter solutions retain an observable trace of the overall quantized character within their fractal profiles.  Finally, the lack of any noticeable form of revival in the FPUT system is, in light of the previous results, not well understood. For this initial value problem, where the energy is uniformly distributed over all wave numbers due to a concentrated initial displacement of a single mass, there is a noticeable disparity between the Korteweg--deVries profiles and those of the FPUT system at such times.

The next stage of this project will be to investigate which of these properties, if any, carry over to the other nonlinear FPUT systems and other nonlinear lattices of interest.  Keeping in mind the numerical observations of dispersive quantization in the Korteweg--deVries equation and its generalizations, \rf{COdisp}, we expect that this will indeed be the case.  Numerical schemes that retain accuracy over long times, while allowing for efficient computation of longer chains, will be essential to this endeavor. This may require the construction of novel numerical methods, e.g., exponential-type Fourier integrators in the spirit of \rf{HoSc}, in addition to the splitting methods considered above.

\vglue 30pt

\Ack  We wish to thank Rajendra Beekie, Gong Chen, Burak Erdo\utxt gan, Natalie Sheils, and Ferdinand Verhulst for helpful discussions and correspondence on FPUT and fractalization. We also thank the referees for all their comments.  Ari Stern was supported in part by a grant from the National Science Foundation (DMS-1913272).

\vskip10pt

\References

\vskip10pt

\key ArKoTe \paper Arioli, G., Koch, H., Terracini, S.; Two novel methods and multi-mode periodic solutions for the Fermi-Pasta-Ulam model; Commun. Math. Phys.; 255 (2005) 1--19\par 
\key BBM \paper Benjamin, T.B., Bona, J.L., Mahony, J.J.; Model equations for long waves in nonlinear dispersive systems; Phil. Trans. Roy. Soc. London A; 272 (1972) 47--78\par
\key Berry \paper Berry, M.V.; Quantum fractals in boxes; J. Phys. A; 29 (1996) 6617--6629\par
\key BerryKlein \paper Berry, M.V., Klein, S.; Integer, fractional and fractal Talbot effects; J. Mod. Optics; 43 (1996) 2139--2164\par
\key BMS \paper Berry, M.V., Marzoli, I., Schleich, W.; Quantum carpets, carpets of light; Physics World; 14{\rm (6)} (2001) 39--44\par
\key BlMo \paper Blanes, S., Moan, P.C.; Practical symplectic partitioned Runge--Kutta and Runge--Kutta--Nystr\"{o}m methods; J. Comput. Appl. Math.; 142 (2002) 313--330\par
\key BonaSaut \paper Bona, J.L., Saut, J.--C.; Dispersive blowup of solutions of generalized Korteweg--de Vries equations; J. Diff. Eq.; 103 (1993) 3--57 \par
\key BOPS \preprint Boulton, L., Olver, P.J., Pelloni, B., Smith, D.A.; New revival phenomena for linear integro-differential equations; {\tt arXiv:2010.01320}, 2020\par
\key Boussinesq \paper Boussinesq, J.; Essai sur la th\'eorie des eaux courantes; M\'em. Acad. Sci. Inst. Nat. France; 23 {\rm (1)} (1877) 1--680\par
\key BruVer \paper Bruggeman, R., Verhulst, F.; Near-integrability and recurrence in FPU chains with alternating masses; J. Nonlinear Sci.; 29 (2019) 183--206\par
\key Calogero \paper Calogero, F.; Solution of the one-dimensional $n$-body problems with quadratic and/or inversely quadratic pair potentials; J. Math. Phys.; 12  (1971) 419--436\par
\key COdisp \paper Chen, G., Olver, P.J.; Dispersion of discontinuous periodic waves; Proc. Roy. Soc. London A; 469 (2013) 20120407\par
\key ChErTz \paper Chousionis, V., Erdo\utxt gan, M.B., Tzirakis, N.; Fractal solutions of linear and nonlinear dispersive partial differential equations; Proc. London Math. Soc.; 110 (2015) 543--564\par
\key DarHua \paper Daripa, P., Hua, W.; A numerical study of an ill-posed Boussinesq equation arising in water waves and nonlinear lattices: Filtering and regularization techniques; Appl. Math. Comput.; 101 (1999) 159--207\par
\key FPUT \paper Dauxois, T.; Fermi, Pasta, Ulam, and a mysterious lady; Physics Today\/; 61{\rm(1)} (2008) 55--57\par
\key DPR \paper Dauxois, T., Peyrard, M., Ruffo, S.; The Fermi-Pasta-Ulam `numerical experiment': history and pedagogical perspectives; European J. Phys.; 26 (2005); S3--S11\par
\key DJ \book Drazin, P.G., Johnson, R.S.; Solitons\/{\rm :} An Introduction; Cambridge University Press, Cambridge, 1989\par
\key DMN \paper Dubrovin, B.A., Matveev, V.B., Novikov, S.P.; Nonlinear equations of Korteweg-de Vries type, finite-zone linear operators and Abelian varieties; Russian Math. Surveys; 31 (1976) 56-134\par
\key ErSh \paper Erdo\utxt gan, M.B., Shakan, G.; Fractal solutions of dispersive partial differential equations on the torus; Selecta Math.; 25 (2019) 11\par
\key ErTznls \paper Erdo\utxt gan, M.B., Tzirakis, N.; Talbot effect for the cubic nonlinear Schr\"odinger equation on the torus; Math. Res. Lett.; 20 (2013) 1081--1090\par
\key ErTz \book Erdo\utxt gan, M.B., Tzirakis, N.; Dispersive Partial Differential Equations: Wellposedness and Applications; London Math. Soc. Student Texts, vol. 86, Cambridge University Press, Cambridge, 2016\par
\key FPU \inbook Fermi, E., Pasta, J., Ulam, S.; Studies of nonlinear problems. I., Los Alamos Report LA1940, 1955; Nonlinear Wave Motion; A.C. Newell, ed., Lectures in Applied Math., vol. 15, American Math. Soc., Providence, R.I., 1974, pp. 143--156\par
\key Ford \paper Ford, J.; The Fermi-Pasta-Ulam problem: Paradox turns discovery; Phys. Rep.; 213 (1992) 271--310\par
\key FrWa \paper Friesecke, G., Wattis, J.; Existence theorem for solitary waves on lattices; Commun. Math. Phys.; 161 (1994) 391--418\par
\key GGMV \paper Galgani, L., Giorgilli, A., Martinoli, A., Vanzini, S.; On the problem of energy partition for large systems of the Fermi--Pasta--Ulam type: analytical and numerical estimates; Physica D; 59 (1992) 334--348\par 
\key HLW \book Hairer, E., Lubich, C., Wanner, G.; Geometric Numerical Integration; Second Edition, Springer--Verlag, New York, 2006\par
\key HoSc \paper Hofmanov\'a, M., Schratz, K.; An exponential-type integrator for the KdV equation; Numer. Math.; 136 (2017) 1117--1137\par
\key LJ \paper Lennard--Jones, J.E.; On the determination of molecular fields; Phys. Rev. A; 5 (1972) 1372--1376\par
\key LPRSV1 \paper Livi, R., Pettini, M., Ruffo, S., Sparpaglione, M., Vulpiani, A.; Relaxation to different stationary states in the Fermi-Pasta-Ulam model; Phys. Rev. A; 28 (1983) 3544--3552\par 
\key LPRSV2 \paper Livi, R., Pettini, M., Ruffo, S., Sparpaglione, M., Vulpiani, A.; Equipartition threshold in nonlinear large Hamiltonian systems: The Fermi--Pasta--Ulam model; Phys. Rev. A; 31 (1985) 1039--1045\par
\key McKvM \paper McKean, H.P., van Moerbeke, P.; The spectrum of Hill's equation; Invent. Math.; 30 (1975) 217--274\par
\key McSt \paper McLachlan, R.I., Stern, A.; Modified trigonometric integrators; SIAM J. Numer. Anal.; 52 (2014) 1378--1397\par
\key McQu \paper McLachlan, R.I., Quispel, G.R.W.; Splitting methods; Acta Numer.; 11 (2002) 341--434\par
\key MoserH \paper Moser, J.; Three integrable Hamiltonian systems connected with isospectral deformations; Adv. Math.; 16 (1975) 197--220\par
\key OHamwwa \paper Olver, P.J.; Hamiltonian perturbation theory and water waves; Contemp. Math.; 28 (1984) 231--249\par
\key OHamwwb \inbook Olver, P.J.; Hamiltonian and non--Hamiltonian models for water waves; Trends and Applications of Pure Mathematics to Mechanics; P.G. Ciarlet and M. Roseau, eds., Lecture Notes in Physics, vol. 195, Springer--Verlag, New York, 1984, pp. 273--290\par
\key Odq \paper Olver, P.J.; Dispersive quantization; Amer. Math. Monthly; 117 (2010) 599--610\par
\key P \book Olver, P.J.; Introduction to Partial Differential Equations; Undergraduate Texts in Mathematics, Springer, New York, 2014\par
\key OS \book Olver, P.J., Shakiban, C.; Applied Linear Algebra; Second Edition, Undergraduate Texts in Mathematics, Springer, New York, 2018\par
\key OTpcon \paper Olver, P.J., Tsatis, E.; Points of constancy of the periodic linearized Korteweg--deVries equation; Proc. Roy. Soc. London A; 474 (2018) 20180160\par
\key OskolkovV \inbook Oskolkov, K.I.; A class of I.M. Vinogradov's series and its applications in harmonic analysis; Progress in Approximation Theory; Springer Ser. Comput. Math., 19, Springer, New York, 1992, pp. 353--402\par
\key Pankov \book Pankov, A.; Travelling Waves and Periodic Oscillations in Fermi--Pasta--Ulam Lattices; Imperial College Press, London, 2005\par 
\key PZHC \paper Porter, M.A., Zabusky, N.J., Hu, B., Campbell, D.K.; Fermi, Pasta, Ulam, and the birth of experimental mathematics; Amer. Scientist\/; 97{\rm(3)} (2009) 214--221\par
\key KAMP \inbook P\"oschel, J.; A lecture on the classical KAM theorem; Proc. Sympos. Pure Math.; 69 (2001) 707--732\par 
\key Rink \paper Rink, B.; Symmetry and resonance in periodic FPU chains; Commun. Math. Phys.; 218 (2001) 665--685\par
\key Rodnianskif \paper Rodnianski, I.; Fractal solutions of the Schr\"odinger equation; Contemp. Math.; 255 (2000) 181--187\par
\key RFPU \paper Rosenau, P.; Dynamics of nonlinear mass-spring chains near the continuum limit; Phys. Lett. A; 118 (1986) 222--227\par
\key RFPUB \paper Rosenau, P.; Dynamics of dense lattices; Phys. Rev. B; 36 (1987) 5868--5876\par
\key Scott \paper Scott, A.C.; Soliton oscillations in DNA; Phys. Rev. A; 31 (1985) 3518--3519\par
\key Smoller \book Smoller, J.; Shock Waves and Reaction--Diffusion Equations; 2nd ed., Springer--Verlag, New York, 1994\par
\key StGr \paper Stern, A., Grinspun, E.; Implicit-explicit variational integration of highly oscillatory problems; Multiscale Model. Simul.; 7 (2009) 1779--1794\par
\key Strang \paper Strang, G.; On the construction and comparison of difference schemes; SIAM J. Numer. Anal.; 5 (1968) 506--517\par
\key Sutherland \paper Sutherland, B.; Exact results for a quantum many-body problem in one-dimension. II; Phys. Rev. A; 5 (1972) 1372--1376\par
\key Talbot \paper Talbot, H.F.; Facts related to optical science. No. IV; Philos. Mag.; 9 (1836) 401--407\par
\key Toda \book Toda, M.; Theory of Nonlinear Lattices; Springer--Verlag, New York, 1981\par
\key TuckMenzel \paper Tuck, J.L., Menzel, M.T.; The superperiod of the nonlinear weighted string (FPU) problem; Adv. Math.; 9 (1972) 399--407\par
\key Ulam \book Ulam, S.M.; Adventures of a Mathematician; Scribner, New York, 1976\par
\key VVS \paper Vrakking, M.J.J., Villeneuve, D.M., Stolow, A.; Observation of fractional revivals of a molecular wavepacket; Phys. Rev. A; 54  (1996) R37--40\par
\key Weissert \book Weissert, T.P.; The Genesis of Simulation in Dynamics: Pursuing the Fermi-Pasta-Ulam Problem; Springer, New York, 1997\par
\key Whitham \book Whitham, G.B.; Linear and Nonlinear Waves; John Wiley \& Sons, New York, 1974\par
\key YeaStr \paper Yeazell, J.A., Stroud, C.R.{, Jr.}; Observation of fractional revivals in the evolution of a Rydberg atomic wave packet; Phys. Rev. A; 43 (1991) 5153--5156\par
\key Zab \paper Zabusky, N.J.; Computational synergetics and mathematical innovation; J. Comp. Phys.; 43 (1981) 195--249\par
\key ZabDeem \paper Zabusky, N.J., Deem, G.S.; Dynamics of nonlinear lattices I. Localized optical excitations, acoustic radiation, and strong nonlinear behavior; J. Comp. Phys.; 2 (1967) 126--153\par
\key ZabKru \paper Zabusky, N.J., Kruskal, M.D.; Interaction of ``solitons'' in a collisionless plasma and the recurrence of initial states; Phys. Rev. Lett.; 15 (1965) 240--243\par

\endRefs

\end{document}